\documentclass[english,pra, twocolumn, superscriptaddress, nofootinbib, showpacs]{revtex4-1}

\usepackage[T1]{fontenc}
\usepackage[latin9]{inputenc}
\usepackage{color}
\usepackage{babel}
\usepackage{amsmath}
\usepackage{amssymb}
\usepackage{graphicx}
\usepackage{wasysym}
\usepackage[unicode=true,pdfusetitle,
 bookmarks=true,bookmarksnumbered=false,bookmarksopen=false,
 breaklinks=false,pdfborder={0 0 1},backref=false,colorlinks=true]
 {hyperref}
\hypersetup{
 citecolor=blue, urlcolor=blue}
\begin{document}

\title{Entropy dynamics of a dephasing model in a squeezed thermal bath}

\author{Yi-Ning You}

\affiliation{Texas A\&M University, College Station, Texas 77843, USA}

\affiliation{University of Science and Technology of China, Hefei, Anhui 230026,
China}

\author{Sheng-Wen Li}

\affiliation{Texas A\&M University, College Station, Texas 77843, USA}
\begin{abstract}
We study the entropy dynamics of a dephasing model, where a two-level
system (TLS) is coupled with a squeezed thermal bath via non-demolition
interaction. This model is exactly solvable, and the time dependent
states of both the TLS and its bath can be obtained exactly. Based
on these states, we calculate the entropy dynamics of both the TLS
and the bath, and find that the dephasing rate of the system relies
on the squeezing phase of the bath. In zero temperature and high temperature
limits, both the system and bath entropy increases monotonically in
the coarse grained time scale. Moreover, we find that the dephasing
rate of the system relies on the squeezing phase of the bath, and
this phase dependence cannot be precisely derived from the Born-Markovian
approximation which is widely adopted in open quantum systems.
\end{abstract}

\maketitle

\section{Introduction}

When an open quantum system is coupled with a thermal reservoir, it
turns out that the classical thermodynamics relations also apply \cite{spohn_entropy_1978,spohn_irreversible_1978,quan_quantum_2005,quan_quantum-classical_2006}.
However, current technology makes it possible to create non-thermal
environment for quantum systems, for example, quantum coherence or
squeezing could also exist in the reservoir, and makes it a non-thermal
bath \cite{scully_extracting_2003,rosnagel_nanoscale_2014,correa_quantum-enhanced_2014,long_performance_2015,agarwalla_quantum_2017,gelbwaser-klimovsky_heat-machine_2014}.
In these cases, it is permissible that the conventional thermodynamics
relations do not hold. Even more strikingly, a quantum heat engine
working with such non-thermal quantum bath could seemingly exceed
the Carnot bound \cite{scully_extracting_2003,rosnagel_nanoscale_2014,correa_quantum-enhanced_2014,agarwalla_quantum_2017}.

This is because indeed conventional thermodynamics only concerns about
thermal equilibrium reservoirs, particularly, the thermal entropy
$dS=\text{\dj}Q/T$ is only defined for equilibrium state. For non-thermal
baths, the conventional entropy relations should be reconsidered.
There are some different approaches dealing with such problems. For
example, some external work should be considered to maintain the quantum
coherence in the bath \cite{quan_quantum_2005,quan_quantum-classical_2006},
excess heat should be taken into account \cite{gardas_thermodynamic_2015},
or the heat should be redefined with the help of passive state \cite{gelbwaser-klimovsky_heat-machine_2014,dag_multiatom_2016}.

Recently, it was noticed that the entropy production in conventional
thermodynamics can be understood as the the correlation generation
between an open quantum system and its thermal reservoir \cite{zhang_general_2008,zhang_information_2009,hilt_system-bath_2009,esposito_entropy_2010,pernice_decoherence_2011,pucci_entropy_2013,manzano_entropy_2016,alipour_correlations_2016,li_production_2017,strasberg_quantum_2017,iyoda_fluctuation_2017}.
For example, for a thermal state $\rho_{\text{\textsc{b}}}(0)=Z^{\text{-}1}\exp[-\hat{H}_{\text{\textsc{b}}}/T]$\footnote{$\hat{H}_{\text{\textsc{s/b}}}$
is the Hamiltonian of the system/bath, and $T$ is the temperature.},
assuming the bath state does not change too much from the initial
state, the von Neumann entropy of the bath state gives $\dot{S}_{\text{\textsc{b}}}=-\mathrm{tr}[\dot{\rho}_{\text{\textsc{b}}}(t)\ln\rho_{\text{\textsc{b}}}(t)]\approx-\mathrm{tr}[\dot{\rho}_{\text{\textsc{b}}}(t)\ln\rho_{\text{\textsc{b}}}(0)]=\frac{d}{dt}\langle\hat{H}_{\text{\textsc{b}}}\rangle/T$
\cite{li_production_2017,scully_radiation_2017}. If all the bath
energy loss is gained by the system, $\delta\langle\hat{H}_{\text{\textsc{b}}}\rangle\approx-\delta\langle\hat{H}_{\text{\textsc{s}}}\rangle$,
then it turns out that the informational entropy change of the bath
$\dot{S}_{\text{\textsc{b}}}$ is just equivalent with the thermal
entropy $\text{\dj}Q/T$ of the system. 

More importantly, in squeezed thermal baths, the conventional thermal
entropy $dS=\text{\dj}Q/T$ does not apply \cite{rosnagel_nanoscale_2014}.
But it turns out that the system-bath (S-B) correlation  still increases
monotonically \cite{li_production_2017}. Therefore, it is worthwhile
to study the entropy dynamics of both the system and its bath in more
examples, especially the exactly solvable models.

In this paper, we study the entropy dynamics in a dephasing model,
where a two-level system (TLS) interacts with a squeezed thermal bath
via non-demolition coupling. This model is well adopted to describe
the physical systems like exciton-phonon interaction, molecular oscillation,
photosynthesis process, etc \cite{dong_photon-blockade_2016,may_charge_2000,dong_how_2017}.
This model is exactly solvable, and plenty of studies have been done
considering the bath is a thermal equilibrium state \cite{breuer_theory_2002,agarwal_quantum_2012,schaller_open_2014}.
Here we consider that the bath starts from a squeezed thermal state
\cite{banerjee_dynamics_2007}, and study the dynamics of both the
system and the bath, especially their entropy. 

We obtain the exact evolution of both the system and bath states,
and find that both the system and bath entropy $S_{\text{\textsc{s,b}}}$
increase monotonically. Moreover, we find that the dephasing rate
of the system relies on the squeezing phase $\theta$ of the bath.
Particularly, in the high temperature limit, the system dephasing
process is Markovian, and the quantum coherence decays exponentially,
with the dephasing rate $\kappa=2\lambda T[\cosh2r-\frac{\ln4}{\pi}\sinh2r\sin\delta\theta]$,
where $r$ is the squeezing strength, $\lambda$ is a unitless number
characterizing the S-B coupling strength, and $\delta\theta$ is the
phase difference between the squeezing phase $\theta$ relative to
the phase of the coupling strength. We also notice that this dephasing
rate $\kappa$ cannot be precisely obtained from the Born-Markovian
approximation widely adopted in open quantum systems.

We arrange the paper as follows: in Sec.\,II, we introduce the dephasing
model, and show how to get the exact evolution operator; in Sec.\,III,
we study the dynamics of the system and its entropy, and discuss the
cases of zero temperature and high temperature limits; in Sec.\,IV,
we study the bath dynamics, and discuss how to calculate the bath
entropy approximately; finally we draw summary in Sec.\,V. Some of
the calculation details are presented in the appendix.

\section{Dephasing model in a squeezed bath \label{sec:Dephasing-model-in}}

Here we first introduce the dephasing model, which is composed of
a TLS ($\hat{H}_{\text{\textsc{s}}}=\frac{1}{2}\Omega_{0}\hat{\sigma}^{z}$)
coupled with a boson bath ($\hat{H}_{\text{\textsc{b}}}=\sum_{k}\omega_{k}\hat{b}_{k}^{\dagger}\hat{b}_{k}$)
\cite{schaller_open_2014,breuer_theory_2002,agarwal_quantum_2012}.
The system-bath interaction is described by the following non-demolition
Hamiltonian,
\begin{equation}
\hat{V}_{\text{\textsc{sb}}}=\hat{\sigma}^{z}\cdot\sum_{k}(g_{k}\hat{b}_{k}+g_{k}^{*}\hat{b}_{k}^{\dagger}),\label{eq:V_SB}
\end{equation}
where $\hat{\sigma}^{z}=|e\rangle\langle e|-|g\rangle\langle g|$,
and $|e\rangle$, $|g\rangle$ are the excited and ground states. 

Notice that $[\hat{H}_{\text{\textsc{s}}},\,\hat{V}_{\text{\textsc{sb}}}]=0$,
thus the system energy is always conserved, and the populations $p_{e,g}$
on $|e,g\rangle$ do not change with time. But the bath energy is
not conserved since $[\hat{H}_{\text{\textsc{b}}},\,\hat{V}_{\text{\textsc{sb}}}]\neq0$.
Therefore, unlike the discussions in conventional thermodynamics,
this open system can never exchange energy with its thermal reservoir
\cite{xu_noncanonical_2014,dong_quantum_2007}. But it is still meaningful
to discuss the information and correlation exchange between the system
and the bath \cite{li_production_2017,dong_how_2017}, as we will
do below.

The evolution behavior of the total S-B system is exactly solvable
\cite{schaller_open_2014,breuer_theory_2002,agarwal_quantum_2012}.
In the interaction picture of $\hat{H}_{\text{\textsc{s}}}+\hat{H}_{\text{\textsc{b}}}$,
it turns out that the evolution operator can be written down as a
separable form $U_{I}(t)=|e\rangle\langle e|\otimes U_{e}+|g\rangle\langle g|\otimes U_{g}$,
where
\begin{gather}
U_{e}=\prod_{k}\hat{\text{\textsc{d}}}_{k}^{+},\qquad U_{g}=\prod_{k}\hat{\text{\textsc{d}}}_{k}^{-},\nonumber \\
\hat{\text{\textsc{d}}}_{k}^{\pm}:=\exp\big\{\pm[\alpha_{k}(t)b_{k}^{\dagger}-\alpha_{k}^{\ast}(t)b_{k}]\big\}.\label{eq:U-evo}
\end{gather}
Notice that $U_{e,g}$ are both products of $\hat{\text{\textsc{d}}}_{k}^{\pm}$,
where $\hat{\text{\textsc{d}}}_{k}^{\pm}$ is a displacement operator
for mode $\hat{b}_{k}$, and we denote 
\begin{equation}
\alpha_{k}(t):=\mu_{k}(1-e^{i\omega_{k}t}),\qquad\mu_{k}:=g_{k}/\omega_{k}
\end{equation}
 for the amount of displacement (see also derivation in Appendix \ref{sec:Evolution-operator}).

With this evolution operator, we can obtain the exact state $\rho_{\text{\textsc{sb}}}(t)$
at any time starting from $\rho_{\text{\textsc{sb}}}(0)=\rho_{\text{\textsc{s}}}(0)\otimes\rho_{\text{\textsc{b}}}(0)$
(hereafter all the density matrices are in the interaction picture).
Plenty of studies have been done considering the initial state of
the bath as a thermal equilibrium state \cite{breuer_theory_2002,morozov_decoherence_2012,marcantoni_thermodynamics_2017}.
In this paper we study the case that the initial state of the boson
bath is a squeezed thermal state \cite{banerjee_dynamics_2007}
\begin{equation}
\rho_{\text{\textsc{b}}}(0)=\hat{{\cal S}}\rho_{\mathrm{th}}\hat{{\cal S}}^{\dagger},\qquad\rho_{\mathrm{th}}=\frac{1}{Z}e^{-\frac{1}{T}\hat{H}_{B}},\label{eq:Sq-bath-state}
\end{equation}
where $T$ is the temperature for the thermal state $\rho_{\mathrm{th}}$,
$Z$ is the normalization constant, $\hat{{\cal S}}=\prod_{k}\hat{\mathfrak{s}}_{k}$
is the squeezing operator for the boson bath, and $\hat{\mathfrak{s}}_{k}$
is the squeezing operator for $\hat{b}_{k}$ mode:
\begin{equation}
\hat{\mathfrak{s}}_{k}=\exp[\frac{1}{2}\xi_{k}^{\ast}\hat{b}_{k}^{2}-\frac{1}{2}\xi_{k}(\hat{b}_{k}^{\dagger})^{2}],\quad\xi_{k}=r_{k}e^{i\theta_{k}}\,(r_{k}\ge0).
\end{equation}
Here $r_{k}$ and $\theta_{k}$ indicates the squeezing strength and
phase respectively. 

\section{System dynamics}

In this section, we study the dynamics of the open system $\rho_{\text{\textsc{s}}}(t)=\mathrm{tr}_{\text{\textsc{b}}}[U_{I}(t)\,\rho_{\text{\textsc{sb}}}(0)\,U_{I}^{\dagger}(t)]$.
Since the populations $p_{e,g}$ on $|e,g\rangle$ do not change with
time, the density matrix of the open system $\rho_{\text{\textsc{s}}}(t)$
can be written as
\begin{equation}
\rho_{\text{\textsc{s}}}(t)=\left[\begin{array}{cc}
p_{e} & \rho_{eg}e^{-\Gamma(t)}\\
\rho_{ge}e^{-\Gamma(t)} & p_{g}
\end{array}\right].\label{eq:rho_S}
\end{equation}
 The time-dependent behavior of the off-diagonal terms shows as
\begin{align}
e^{-\Gamma(t)} & =\mathrm{tr}_{\text{\textsc{b}}}[U_{e}\,\rho_{\text{\textsc{b}}}(0)\,U_{g}^{\dagger}]\nonumber \\
= & \mathrm{tr}_{\text{\textsc{b}}}\Big(\rho_{\text{\textsc{b}}}(0)\,\exp\big\{2\sum_{k}[\alpha_{k}(t)\hat{b}_{k}^{\dagger}-\alpha_{k}^{*}(t)\hat{b}_{k}]\big\}\Big).
\end{align}
If the decay factor $\Gamma(t)$ linearly depends on $t$, it means
the quantum coherence terms decay exponentially and that is a Markovian
process \cite{breuer_theory_2002,gardiner_quantum_2004,chruscinski_markovianity_2012}. 

Notice that the above expression for $e^{-\Gamma(t)}$ is just the
characteristic function for the Wigner representation of $\rho_{\text{\textsc{b}}}(0)$,
which is a squeezed thermal state of all bath modes \cite{scully_quantum_1997,gardiner_quantum_2004,agarwal_quantum_2012,li_long-term_2014}.
Thus we obtain
\begin{equation}
\Gamma(t)=\sum_{k}\frac{1}{2}|\gamma_{k}(t)|^{2}\coth\frac{\omega_{k}}{2T},
\end{equation}
where $\gamma_{k}(t):=2\alpha_{k}(t)\cosh r_{k}+2\alpha_{k}^{*}(t)e^{i\theta_{k}}\sinh r_{k}$.
Substituting $\alpha_{k}(t):=\frac{g_{k}}{\omega_{k}}(1-e^{i\omega_{k}t})$
into the above expression, the decay factor becomes \begin{widetext}
\begin{equation}
\Gamma(t)=\sum_{k}\frac{4|g_{k}|^{2}}{\omega_{k}^{2}}(1-\cos\omega_{k}t)\coth\frac{\omega_{k}}{2T}\big[\cosh2r_{k}-\sinh2r_{k}\cos(\omega_{k}t-\Delta\theta_{k})\big],
\end{equation}
where $\Delta\theta_{k}:=\theta_{k}-2\phi_{k}$ is the phase difference
between the squeezing phase $\theta_{k}$ relative to the phase of
the coupling strength $g_{k}$ ($g_{k}=|g_{k}|e^{i\phi_{k}}$, $\phi_{k}:=\arg[g_{k}]$).

Now we introduce a coupling spectral density $J(\omega):=2\pi\sum_{k}|g_{k}|^{2}\,\delta(\omega-\omega_{k})$
\cite{breuer_theory_2002,li_steady_2015}, then the above summation
can be written as an integral for continuous bath modes (considering
$r_{k}=r$, $\Delta\theta_{k}=\delta\theta$ are constants):
\begin{equation}
\Gamma(t)=\int_{0}^{\infty}\frac{d\omega}{2\pi}\,4J(\omega)\coth\frac{\omega}{2T}\cdot\frac{1-\cos\omega t}{\omega^{2}}\big[\cosh2r-\sinh2r\cos(\omega t-\delta\theta)\big].\label{eq:integral}
\end{equation}
\end{widetext}

Here we adopt the Ohmic coupling spectral density with an exponential
cutoff (with cutoff frequency $\Omega_{c}$), i.e., $J(\omega)=\lambda\omega\,e^{-\omega/\Omega_{c}}$,
which could leads to Markovian process in many cases \cite{breuer_theory_2002,gardiner_quantum_2004,li_non-markovianity_2016}.
Here $\lambda$ is a unitless number indicating the coupling strength.

\subsection{Zero temperature case \label{subsec:Zero-temperature-case}}

When the temperature $T\rightarrow0$, we have $\coth\frac{\omega}{2T}\rightarrow1$.
In this case, the initial state of the bath is a pure state squeezed
from the vacuum. The decay factor of the open system can be integrated
out, and that is
\begin{equation}
\Gamma(t)=\frac{\lambda}{\pi}\big[\mathsf{A}_{t}\cosh2r-\sinh2r\,(\mathsf{B}_{t}\cos\delta\theta+\mathsf{C}_{t}\sin\delta\theta)\big],\label{eq:Gam(t)}
\end{equation}
where we denote ($\tau:=\Omega_{c}t$)
\begin{gather}
\mathsf{A}_{t}=\ln[1+\tau^{2}],\quad\mathsf{B}_{t}=\ln\frac{[1+4\tau^{2}]^{\frac{1}{2}}}{1+\tau^{2}},\nonumber \\
\mathsf{C}_{t}=2\tan^{\text{-1}}\tau-\tan^{\text{-1}}2\tau.\label{eq:ABC-0}
\end{gather}

It is simple to see that $\mathsf{A}_{t}$ and $\mathsf{B}_{t}$ both
give rise to a power-law decay behavior. However, the factor $\mathsf{C}_{t}$
shows a quite different decay behavior. For very short time scale
$t\apprle\Omega_{c}^{\text{-}1}$, we have $\mathsf{C}_{t}\thickapprox2\Omega_{c}^{3}t^{3}$,
which leads to a cubic exponential decay behavior $\sim\exp[-\frac{2\lambda}{\pi}\Omega_{c}^{3}t^{3}]$.
But for long time scale $t\gg\Omega_{c}^{\text{-}1}$ ($\Omega_{c}t\gg1$),
approximately we have $\tan^{\text{-1}}\Omega_{c}t\thickapprox\tan^{\text{-1}}2\Omega_{c}t\thickapprox\pi/2$,
and thus $\mathsf{C}_{t}\approx\pi/2$ is just a constant. This is
quite different from the results of thermal baths \cite{breuer_theory_2002}.

\begin{figure}
\includegraphics[width=1\columnwidth]{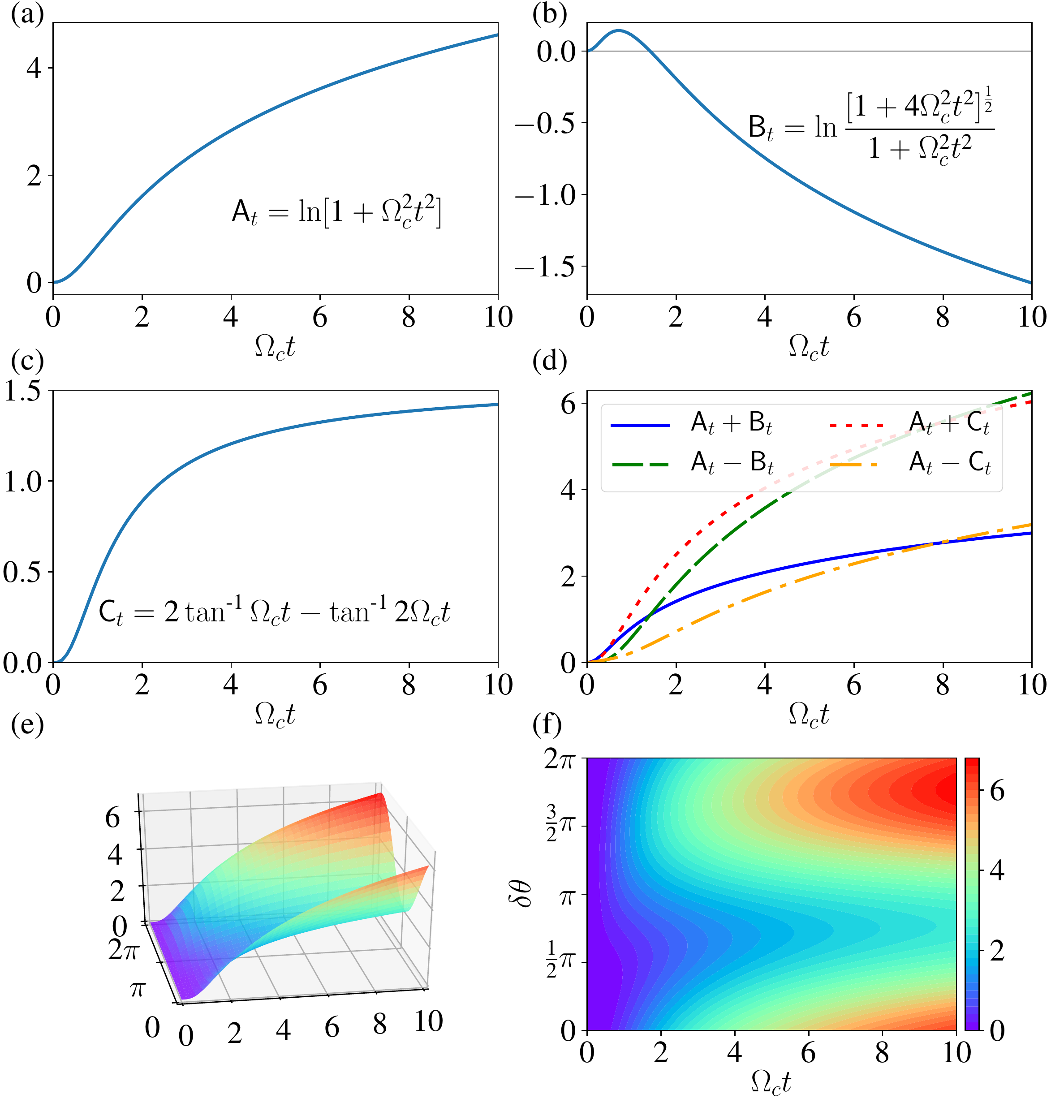}\caption{(Color online) The time-dependence of the coefficients in zero temperature
{[}Eq.\,(\ref{eq:ABC-0}){]} (a) $\mathsf{A}_{t}$, (b) $\mathsf{B}_{t}$,
(c) $\mathsf{C}_{t}$, (d) $\mathsf{A}_{t}\pm\mathsf{B}_{t}$, $\mathsf{A}_{t}\pm\mathsf{C}_{t}$,
and (e, f) $\mathsf{A}_{t}-(\mathsf{B}_{t}\cos\delta\theta+\mathsf{C}_{t}\sin\delta\theta)$.}

\label{fig-0ABC}
\end{figure}

We show the time-dependent coefficients $\mathsf{A}_{t}$, $\mathsf{B}_{t}$
and $\mathsf{C}_{t}$ in Fig.\,\ref{fig-0ABC}. By checking the positivity
of $d\Gamma(t)/dt$ in the area $t\ge0$, it is straightforward to
prove that $\Gamma(t)$ is a monotonically increasing function for
any squeezing parameters $r$ and $\theta$, which means the coherence
of the TLS always decays monotonically (see Appendix \ref{sec:The-increasing-rate}).

\subsection{High temperature limit}

Now we consider the high temperature limit. In this case, we have
$\coth\frac{\omega}{2T}\thickapprox\frac{2T}{\omega}$, and put it
into the integral Eq.\,(\ref{eq:integral}). However, the singularity
in the denominator ($\omega^{2}$) still makes it uneasy for the integration.
Here we eliminate this singularity by taking the derivate of $t$
to the 2nd order in the integral \cite{breuer_theory_2002}, and it
turns out that $\partial_{t}^{2}\Gamma(t)$ can be integrated out.
Then the decay factor $\Gamma(t)$ can be obtained by integrating
over $t$ under the initial conditions $\Gamma|_{t=0}$ and $\partial_{t}\Gamma|_{t=0}$.
Obviously, from Eq.\,(\ref{eq:rho_S}) we know $\Gamma|_{t=0}=0$.
And $\partial_{t}\Gamma|_{t=0}$ can be obtained by the integration
of Eq.\,(\ref{eq:integral}) by taking the derivative of $t$ to
the 1st order and then setting $t=0$, which gives $\partial_{t}\Gamma|_{t=0}=0$.
It turns out that the decay factor $\Gamma(t)$ still has the same
form as Eq.\,(\ref{eq:Gam(t)}), but the time-dependent coefficients
$\mathsf{A}_{t}$, $\mathsf{B}_{t}$ and $\mathsf{C}_{t}$ now become
($\tau:=\Omega_{c}t$)
\begin{align}
\mathsf{A}_{t} & =\frac{2T}{\Omega_{c}}\Big(2\tau\tan^{\text{-1}}\tau-\ln[1+\tau^{2}]\Big),\label{eq:ABC-high}\\
\mathsf{B}_{t} & =\frac{2T}{\Omega_{c}}\Big(2\tau[\tan^{\text{-1}}2\tau-\tan^{\text{-1}}\tau]-\ln\frac{[1+4\tau^{2}]^{\frac{1}{2}}}{1+\tau^{2}}\Big),\nonumber \\
\mathsf{C}_{t} & =\frac{2T}{\Omega_{c}}\Big([\tan^{\text{-1}}2\tau-2\tan^{\text{-1}}\tau]+\tau\ln\frac{1+4\tau^{2}}{1+\tau^{2}}\Big).\nonumber 
\end{align}

For the time scale $t\gg\Omega_{c}^{\text{-}1}$, considering $\Omega_{c}\gg T$,
the above time-dependent factors become
\begin{equation}
\mathsf{A}_{t}\thickapprox2\pi Tt,\quad\mathsf{B}_{t}\thickapprox0,\quad\mathsf{C}_{t}\thickapprox2Tt\,\ln4.\label{eq:kappa}
\end{equation}
Therefore, the decay factor $\Gamma(t)$ depends linearly on $t$,
$\Gamma(t)=\kappa t$, where we define the decay rate as 
\begin{equation}
\kappa:=2\lambda T[\cosh2r-\frac{\ln4}{\pi}\sinh2r\sin\delta\theta].\label{eq:dephase}
\end{equation}
 That means, the coherence of the TLS decays exponentially, and this
is a Markovian process. Notice that $\Omega_{c}^{\text{-}1}$ is a
very short time comparing with the system dynamics, and $t\gg\Omega_{c}^{\text{-}1}$
just means after the relaxation time of the bath. When there is no
squeezing, this result returns to the thermal bath result in previous
studies \cite{breuer_theory_2002,marcantoni_thermodynamics_2017}.

Notice that here $\ln4/\pi\thickapprox0.44<1$, thus the decay rate
$\kappa$ is always positive for any phase difference $\delta\theta$.
It is worth noticing that the decay rate $\kappa$ depends on the
phase difference $\delta\theta$. Especially, when $\delta\theta=3\pi/2$,
we have $\sin\delta\theta=-1$, thus the decay rate $\kappa$ is suppressed;
also, when $\delta\theta=\pi/2$, we have $\sin\delta\theta=1$, and
the decay rate $\kappa$ gets the maximum enhancement. When the squeezing
strength $r$ is strong, $\cosh r\thickapprox\sinh r\thickapprox e^{r}/2$,
thus the decay rate is $\kappa\thickapprox\lambda Te^{r}[1-\sin\delta\theta(\ln4/\pi)]$.

Using the Born-Markovian approximation \cite{breuer_theory_2002},
we can also derive a master equation describing the Markovian dephasing
behavior (Appendix \ref{sec:Markovian}), i.e.,
\begin{equation}
\dot{\rho}_{\text{\textsc{s}}}=\frac{1}{2}\kappa'\Big([\sigma_{z}\rho_{\text{\textsc{s}}},\,\sigma_{z}]+[\sigma_{z},\,\rho_{\text{\textsc{s}}}\sigma_{z}]\Big).\label{eq:ME}
\end{equation}
 But the decay rate is $\kappa'=2\lambda T(\cosh2r-\sinh2r\cos\delta\theta).$
For thermal bath case ($r=0$), this Born-Markovian dephasing rate
$\kappa'$ coincides with the one obtain from the exact evolution
\cite{breuer_theory_2002}; but for a squeezed thermal bath, the Born-Markovian
master equation is not precise enough. We will discuss the reason
for this inconsistency later.

\subsection{Exact result and the system entropy}

The exact result can be obtained by making the following expansion
in the integral Eq.\,(\ref{eq:integral}),
\begin{equation}
\coth\frac{\omega}{2T}=1+2\sum_{n=1}^{\infty}e^{-\frac{n\omega}{2T}}.
\end{equation}
The 0-order leads to the same integral as the above zero temperature
case, and the other terms give rise to similar integrals as the above
high temperature case, except the exponential cutoff should be corrected
to be $\exp[-\frac{\omega}{\Omega_{c}}-\frac{n\omega}{2T}]$.

As the result, the decay factor still has the form of Eq.\,(\ref{eq:Gam(t)}),
but the coefficients $\mathsf{A}_{t}$, $\mathsf{B}_{t}$ and $\mathsf{C}_{t}$
are changed to be
\begin{gather}
\mathsf{A}_{t}=\mathsf{a}_{0}+2\sum_{n=1}^{\infty}\mathsf{a}_{n},\qquad\mathsf{B}_{t}=\mathsf{b}_{0}+2\sum_{n=1}^{\infty}\mathsf{b}_{n},\nonumber \\
\mathsf{C}_{t}=\mathsf{c}_{0}+2\sum_{n=1}^{\infty}\mathsf{c}_{n},
\end{gather}
 where $\mathsf{a}_{0}$, $\mathsf{b}_{0}$ and $\mathsf{c}_{0}$
are exactly the same with the coefficients $\mathsf{A}_{t}$, $\mathsf{B}_{t}$
and $\mathsf{C}_{t}$ in Eq.\,(\ref{eq:ABC-0}) (zero temperature
result), and $\mathsf{a}_{n}$, $\mathsf{b}_{n}$ and $\mathsf{c}_{n}$
has the same form with the coefficients $\mathsf{A}_{t}$, $\mathsf{B}_{t}$
and $\mathsf{C}_{t}$ in Eq.\,(\ref{eq:ABC-high}) (high temperature
result), except the cutoff energy $\Omega_{c}$ in Eq.\,(\ref{eq:ABC-high})
should be corrected to be $\Omega_{c}\rightarrow\Omega_{c}\big/[1+\frac{n\Omega_{c}}{2T}]$
correspondingly.

When the off-diagonal terms of the TLS decrease, the system entropy
always increases. The TLS state can be always written as $\rho_{\text{\textsc{s}}}=\frac{1}{2}(\mathbf{1}+\sum v_{i}\hat{\sigma}^{i})$,
where $i=x,y,z$, and $v_{i}:=\mathrm{tr}[\rho_{\text{\textsc{s}}}\hat{\sigma}^{i}]$\@.
Then the two eigenvalues of $\rho_{\text{\textsc{s}}}$ are $\frac{1}{2}(1\pm u)$,
where $u=[v_{x}^{2}+v_{y}^{2}+v_{z}^{2}]^{\frac{1}{2}}\in[0,1]$,
thus the entropy of the TLS is
\begin{gather}
S_{\text{\textsc{s}}}=\ln2-\frac{1}{2}\big[(1+u)\ln(1+u)+(1-u)\ln(1-u)\big],\nonumber \\
\dot{S}_{\text{\textsc{s}}}=-\frac{1}{2}\dot{u}\ln\frac{1+u}{1-u}.
\end{gather}
Notice that in this dephasing model, $v_{z}$ does not change, thus
$\dot{u}=(v_{x}\dot{v}_{x}+v_{y}\dot{v}_{y})/u$ \cite{morozov_decoherence_2012}.

Therefore, when $u$ decreases, the system entropy $S_{\text{\textsc{s}}}$
increases. In both cases of zero temperature and high temperature
limit, the off-diagonal terms decay monotonically to zero, which indicates
$u$ decreases and the system entropy $S_{\text{\textsc{s}}}$ increases
monotonically. 

\section{Bath dynamics}

In this section, we study the dynamics of the bath state, especially
the entropy change of the bath.

Using the evolution operator $U_{I}(t)$ given in Sec.\,\ref{sec:Dephasing-model-in},
we can exactly write down the time-dependent bath state, i.e., 
\begin{equation}
\rho_{\text{\textsc{b}}}(t)=p_{e}\rho_{\text{\textsc{b}}}^{+}(t)+p_{g}\rho_{\text{\textsc{b}}}^{-}(t),\label{eq:bath-evo}
\end{equation}
where $\rho_{\text{\textsc{b}}}^{+}(t)=U_{e}\,\rho_{\text{\textsc{b}}}(0)\,U_{e}^{\dagger}$
and $\rho_{\text{\textsc{b}}}^{-}(t)=U_{g}\,\rho_{\text{\textsc{b}}}(0)\,U_{g}^{\dagger}$.

Notice that the evolution operators $U_{e,g}$ {[}Eq.\,(\ref{eq:U-evo}){]}
are both products of the displacement operators $\hat{\text{\textsc{d}}}_{k}^{\pm}$
of each bath mode $\hat{b}_{k}$. Therefore, similar like the initial
state $\rho_{\text{\textsc{b}}}(0)$, $\rho_{\text{\textsc{b}}}(t)$
always keeps a product form $\rho_{\text{\textsc{b}}}(t)=\bigotimes_{k}\varrho_{k}(t)$,
where 
\begin{equation}
\varrho_{k}(t)=p_{e}\varrho_{k}^{+}(t)+p_{g}\varrho_{k}^{-}(t)
\end{equation}
 is the state of the bath mode $\hat{b}_{k}$, and $\varrho_{k}^{\pm}(t)=\hat{\text{\textsc{d}}}_{k}^{\pm}\varrho_{k}(0)[\hat{\text{\textsc{d}}}_{k}^{\pm}]^{\dagger}$.
Thus the von Neumann entropy of $\rho_{\text{\textsc{b}}}(t)$ can
be calculated by $\dot{S}[\rho_{\text{\textsc{b}}}(t)]=\sum_{k}\dot{S}[\varrho_{k}(t)]$.

The evolution of $\varrho_{k}^{\pm}(t)$ has a quite clear picture
in the phase space of Wigner function, i.e., they are displaced Gaussian
packages with displacement $\pm\alpha_{k}(t)$. Initially, $\varrho_{k}(0)$
is a squeezed thermal state centered at the original point. The operators
$U_{e,g}$ displace $\varrho_{k}(0)$ to the new centers at $\langle\hat{b}_{k}\rangle=\pm\alpha_{k}(t)=\pm\mu_{k}(1-e^{i\omega_{k}t})$
correspondingly. With the time goes by, the package trajectories of
$\varrho_{k}^{\pm}(t)$ form two cycles, and the state $\varrho_{k}(t)$
is their probabilistic mixture. That also means, indeed the bath never
reaches any steady state.

As the result, the state $\varrho_{k}(t)$ of each bath mode is a
non-Gaussian state. Thus it is still difficult to get the analytical
result of its von Neumann entropy, although we know exactly the density
matrix $\varrho_{k}(t)$ \cite{agarwal_entropy_1971,braunstein_quantum_2005,genoni_quantifying_2008}.
To bypass this difficulty, we calculate the dynamics of the bath state
entropy with the following approximation \cite{aurell_von_2015,li_production_2017}:
\begin{equation}
\dot{S}_{\text{\textsc{b}}}=-\mathrm{tr}[\dot{\rho}_{\text{\textsc{b}}}(t)\ln\rho_{\text{\textsc{b}}}(t)]\thickapprox-\mathrm{tr}[\dot{\rho}_{\text{\textsc{b}}}(t)\ln\rho_{\text{\textsc{b}}}(0)],\label{eq:entropy}
\end{equation}
assuming that the the state $\rho_{\text{\textsc{b}}}(t)$ does not
change too much from $\rho_{\text{\textsc{b}}}(0)$ and thus $\ln\rho_{\text{\textsc{b}}}(t)\thickapprox\ln\rho_{\text{\textsc{b}}}(0)$.
This is quite similar with the idea of the Born approximation widely
adopted in open quantum systems \cite{breuer_theory_2002}.

\begin{figure}
\includegraphics[width=1\columnwidth]{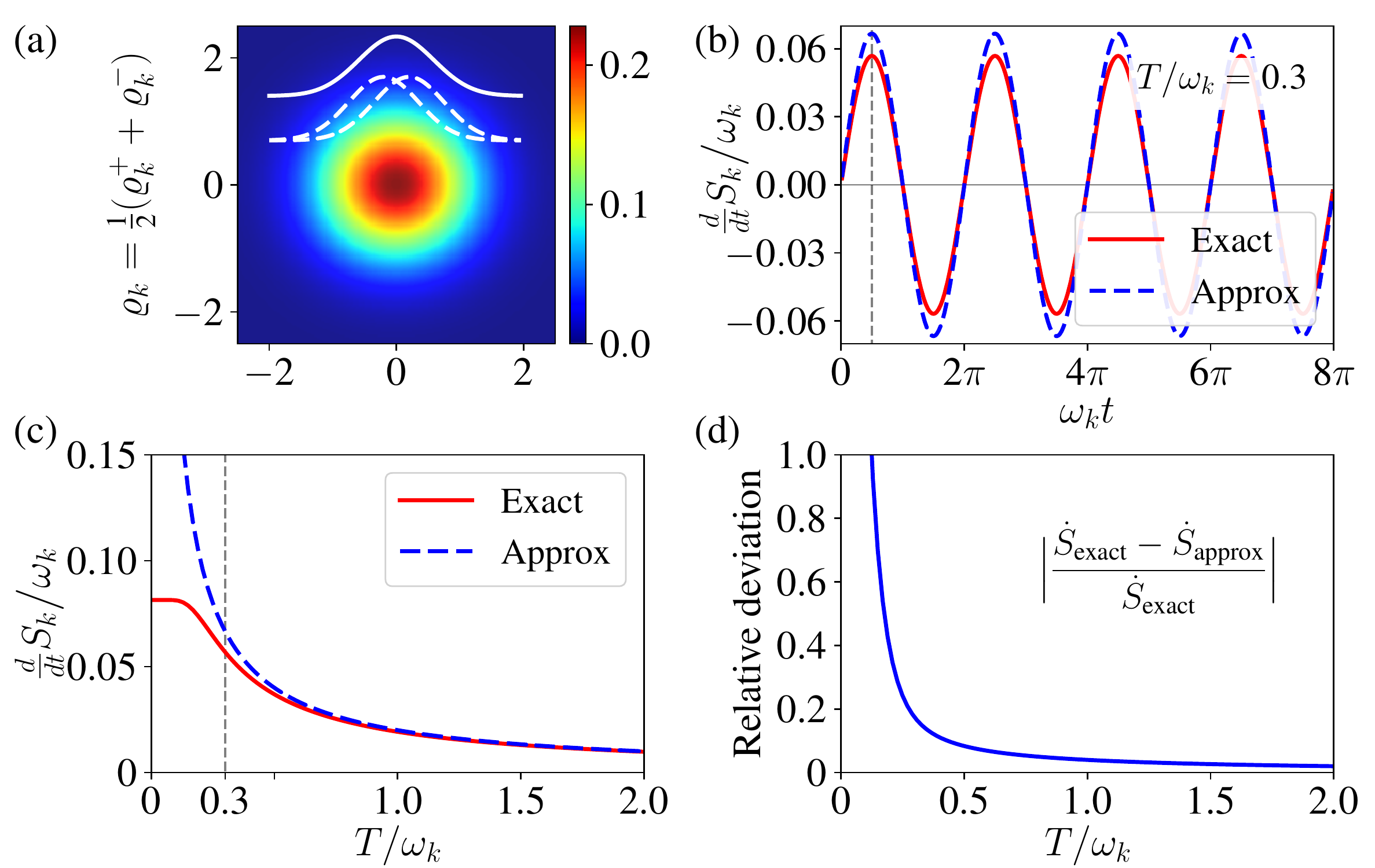}

\caption{(Color online) (a) The Wigner function of a $\varrho_{k}=\frac{1}{2}(\varrho_{k}^{+}+\varrho_{k}^{-})$,
where $\varrho_{k}^{\pm}=\hat{\text{\textsc{d}}}_{k}^{\pm}\varrho_{\mathrm{th}}[\hat{\text{\textsc{d}}}_{k}^{\pm}]^{\dagger}$
are displaced thermal state with displacement $\pm\alpha_{k}(t)=\pm\mu_{k}(1-e^{i\omega_{k}t})$
and $\mu_{k}=0.1$. (b) The time-dependence of $\dot{S}[\varrho_{k}(t)]$
calculated by exact diagonalization (red solid) and the approximation
Eq.\,(\ref{eq:approx}) (blue dashed) for $T/\omega_{k}=0.3$. (c)
Comparison of the exact and approximated results for $\dot{S}[\varrho_{k}]|_{\omega_{k}t=\frac{\pi}{2}}$
at different temperatures. (d) The relative deviation $|1-\dot{S}_{\text{approx}}/\dot{S}_{\text{exact}}|$.}

\label{fig-error}
\end{figure}

Under this approximation, for a thermal bath state $\rho_{\mathrm{th}}(0)=Z^{\text{-1}}\exp(-\hat{H}_{\text{\textsc{b}}}/T)$,
we obtain $\dot{S}_{\text{\textsc{b}}}\thickapprox\frac{d}{dt}\langle\hat{H}_{\text{\textsc{b}}}\rangle/T$,
which has the same form with the thermal entropy $dS=\text{\dj}Q/T$
in conventional thermodynamics \cite{kondepudi_modern_2014}. Similarly,
for a squeezed thermal bath, the bath entropy gives $\dot{S}_{\text{\textsc{b}}}\thickapprox\frac{d}{dt}\langle\hat{{\cal S}}\hat{H}_{\text{\textsc{b}}}\hat{{\cal S}}^{\dagger}\rangle/T$
\cite{rosnagel_nanoscale_2014,li_production_2017}. 

Thus, now the problem of the bath entropy dynamics is converted into
calculating the dynamical variable expectations of the bath \cite{li_production_2017}.
Since the exact evolution of the bath state $\rho_{\text{\textsc{b}}}(t)$
is known ({[}Eq.\,(\ref{eq:bath-evo}){]}), we make a numerical comparison
(Fig.\,\ref{fig-error}) for the above approximation with the result
calculated by exact diagonalization of the density matrix {[}for the
single mode state $\varrho_{k}(t)=p_{e}\varrho_{k}^{+}(t)+p_{g}\varrho_{k}^{-}(t)${]}. 

When the two Gaussian packages $\varrho_{k}^{\pm}(t)$ are quite close
to each other, their mixture $\varrho_{k}(t)$ well looks like a single
Gaussian package {[}Fig.\,\ref{fig-error}(a){]}. The separation
of $\varrho_{k}^{\pm}(t)$ is $2|\alpha_{k}(t)|=\frac{2|g_{k}|}{\omega_{k}}\sqrt{2-2\cos\omega_{k}t}\le4|g_{k}|/\omega_{k}$,
which is determined by the S-B interaction strength $|g_{k}|$, thus
a weaker S-B interaction ($\lambda\ll1$) makes better approximation. 

In Fig.\,\,\ref{fig-error}, we consider an example of a thermal
bath state, and the above approximation gives 
\begin{align}
\dot{S}[\varrho_{k}] & \approx\frac{\omega_{k}}{T}\frac{d}{dt}\langle\hat{b}_{k}^{\dagger}\hat{b}_{k}\rangle=\frac{\omega_{k}}{T}\frac{d}{dt}|\alpha_{k}(t)|^{2}\nonumber \\
 & =\frac{\omega_{k}}{T}\cdot2|\mu_{k}|^{2}\omega_{k}\sin\omega_{k}t=\frac{2|g_{k}|^{2}}{T}\sin\omega_{k}t.\label{eq:approx}
\end{align}
This approximated result has the same oscillation behavior with the
exact one {[}Fig.\,\ref{fig-error}(b){]}, and the amplitudes also
fit well (for $T/\omega_{k}=0.3$). Thus we can use the maximum value
(at $\omega_{k}t=\pi/2$) to characterize the their deviation at different
temperatures.

In Fig.\,\ref{fig-error}(c) we show both the exact and approximated
result for $\dot{S}[\varrho_{k}]|_{\omega_{k}t=\frac{\pi}{2}}$ at
different temperatures, as well as their relative deviation in Fig.\,\ref{fig-error}(d).
It turns out that indeed this approximation works well in the high
temperature regime, but not so well when $T\rightarrow0$. Indeed,
in the approximation (\ref{eq:approx}), it is explicit to see that
$\dot{S}[\varrho_{k}]$ diverges at low temperature, and this similar
divergence behavior also appears in the conventional thermal entropy
$\text{\dj}Q/T$ \cite{santos_wigner_2017}. But the exact result
for the von Neumann entropy $\dot{S}[\varrho_{k}]$ does not diverge
at low temperature {[}red solid line in Fig.\,\ref{fig-error}(c){]}.

This is because the bath entropy $S[\varrho_{k}(t)]$ comes from two
origins: one is the uncertainty due to finite temperature, the other
one comes from the mixture proportion of $\varrho_{k}^{\pm}(t)$ encoded
in the initial state probabilities $p_{e,g}$. From Eq.\,(\ref{eq:approx}),
we see that this approximated result does not depends on the probabilities
$p_{e,g}$, which means this part of uncertainty is omitted, and only
the thermal fluctuation is counted.

In high temperature regime, the entropy of thermal fluctuation dominates
thus the approximation works well; In the low temperature regime,
the thermal uncertainty approaches zero, thus the non-Gaussian property
of $\varrho_{k}(t)$ becomes important, and the approximation is not
good. Therefore, the validity of the approximation (\ref{eq:entropy})
replies on specific model and conditions. In this dephasing model,
the TLS brings in nonlinearity to the model, which gives rise to the
non-Gaussian property of the bath state. In this case, the above approximation
does not work well. 

Here we emphasize that $\dot{S}_{\text{\textsc{b}}}\approx\frac{d}{dt}\langle\hat{H}_{\text{\textsc{b}}}\rangle/T$
describes the bath entropy dynamics, although it has an analogous
form with the thermal entropy $dS=\text{\dj}Q/T$, which is defined
for the open system but not the bath. Thus, the above failure of the
approximation in the low temperature regime is not in conflict with
the conventional thermodynamics.

For the squeezed thermal bath case, in the high temperature regime,
the above ``semi-Born'' approximation gives the entropy of one bath
mode as
\begin{align}
\dot{S}[\varrho_{k} & (t)]\approx\frac{\omega_{k}}{T}\frac{d}{dt}\langle\hat{\mathfrak{s}}_{k}\,\hat{b}_{k}^{\dagger}\hat{b}_{k}\,\hat{\mathfrak{s}}_{k}^{\dagger}\rangle\nonumber \\
= & \frac{2|g_{k}|^{2}}{T}\Big\{\cosh2r_{k}\sin\omega_{k}t-\sinh2r_{k}\times\nonumber \\
 & \qquad\big[\sin(2\omega_{k}t-\Delta\theta_{k})-\sin(\omega_{k}t-\Delta\theta_{k})\big]\Big\},
\end{align}
where $\Delta\theta_{k}=\theta_{k}-2\phi_{k}$, and $\phi_{k}=\arg[g_{k}]$
is the phase of the coupling strength $g_{k}$. It is worth noticing
that, similar as the system dynamics {[}Eq.\,(\ref{eq:dephase}){]},
the entropy changing rate $\dot{S}[\varrho_{k}]$ depends on the squeezing
phase $\Delta\theta_{k}$ of the bath mode.

\begin{figure}
\includegraphics[width=1\columnwidth]{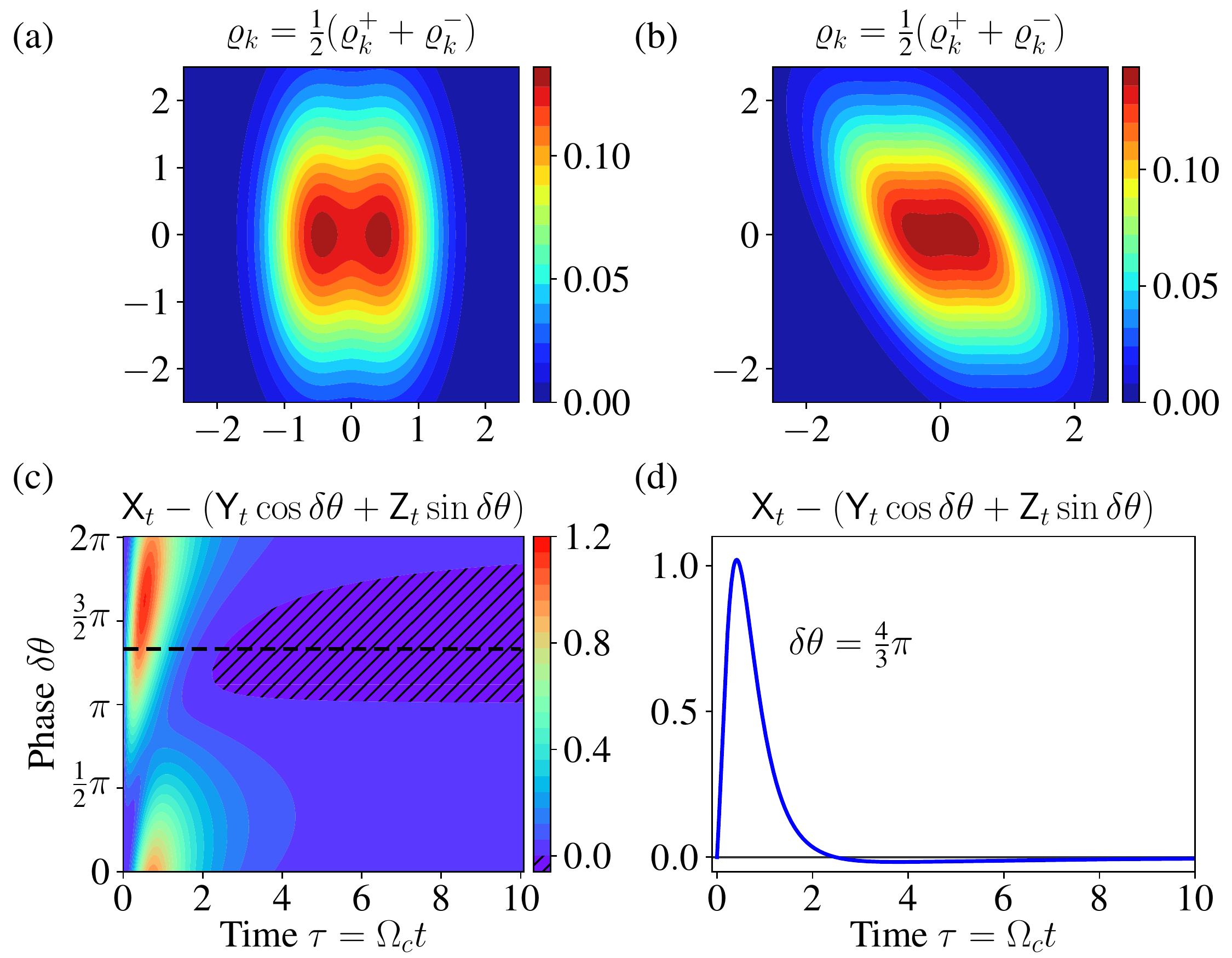}\caption{(Color online) (a, b) The Wigner function for $\varrho_{k}=\frac{1}{2}(\varrho_{k}^{+}+\varrho_{k}^{-})$
at the maximum separation of $\varrho_{k}^{\pm}$ (when $\omega_{k}t=\pi/2$).
The squeezing phases $\theta$ are different in these two figures.
(c, d) $\mathsf{X}_{t}-(\mathsf{Y}_{t}\cos\delta\theta+\mathsf{Z}_{t}\sin\delta\theta)$.
The shadowed area indicates the bath entropy could decrease in this
area, but the decreasing rate is quite small.}

\label{fig-squ}
\end{figure}

This can be intuitively understood from Fig.\,\ref{fig-squ}(a, b),
where the Wigner functions of $\varrho_{k}(t)=p_{e}\varrho_{k}^{+}+p_{g}\varrho_{k}^{-}$
with different squeezing phases are shown at the maximum separation
of $\varrho_{k}^{\pm}$ ($\omega_{k}t=\pi/2$). Obviously, due to
the different squeezing phases, the states $\varrho_{k}(t)$ differ
a lot, and this phase dependence is also reflected in the expectation
values $\langle\hat{b}_{k}^{\dagger}\hat{b}_{k}\rangle$, $\langle\hat{b}_{k}^{2}\rangle$,
and the entropy change $\dot{S}[\varrho_{k}(t)]$. Besides, it is
clear to see that this difference cannot be eliminated by doing any
phase rotation on the initial state.

This also explains why the dephasing rate of the system depends on
the squeezing phase of the bath {[}Eq.\,(\ref{eq:dephase}){]}. Since
only the initial state of the bath is concerned when deriving the
Born-Markovian master equation (Appendix \ref{sec:Markovian}), the
above bath dynamics is not taken into consideration, therefore, the
dephasing rate $\kappa'$ {[}Eq.\,(\ref{eq:ME}){]} derived from
the Born-Markovian approximation does not coincides with the one obtained
from the exact evolution. 

Summing up $\dot{S}[\varrho_{k}]$ for all bath modes, the entropy
changing rate of the total bath is given by the following integral:\begin{widetext}
\begin{equation}
\dot{S}_{\text{\textsc{b}}}=\sum_{k}\dot{S}[\varrho_{k}]=\frac{2}{T}\int_{0}^{\infty}\frac{d\omega}{2\pi}\,J(\omega)\Big\{\cosh2r\sin\omega t-\sinh2r\big[\sin(2\omega t-\delta\theta)-\sin(\omega t-\delta\theta)\big]\Big\}.
\end{equation}
\end{widetext} If we choose the Ohmic coupling spectral density $J(\omega)=\lambda\omega\,e^{-\omega/\Omega_{c}}$
as before, the above integral gives 
\[
\dot{S}_{\text{\textsc{b}}}=\frac{\lambda\Omega_{c}^{2}}{\pi T}\big[\mathsf{X}_{t}\cosh2r-\sinh2r\,(\mathsf{Y}_{t}\cos\delta\theta+\mathsf{Z}_{t}\sin\delta\theta)\big],
\]
 where the coefficients are (denoting $\tau:=\Omega_{c}t$)
\begin{gather}
\mathsf{X}_{t}=\frac{2\tau}{(1+\tau^{2})^{2}},\quad\mathsf{Y}_{t}=\frac{2\tau(1-4\tau^{2}-14\tau^{4})}{(1+\tau^{2})^{2}(1+4\tau^{2})^{2}},\nonumber \\
\mathsf{Z}_{t}=\frac{3\tau^{2}(3+5\tau^{2}-4\tau^{4})}{(1+\tau^{2})^{2}(1+4\tau^{2})^{2}}.
\end{gather}

When there is no squeezing in the bath ($r=0$), $\dot{S}_{\text{\textsc{b}}}=\frac{\lambda\Omega_{c}^{2}}{\pi T}\mathsf{X}_{t}$,
which is always positive for $t>0$, meaning the bath entropy increases
monotonically. The terms with $\mathsf{Y}_{t}$ and $\mathsf{Z}_{t}$
can either increase or decrease with time, depending on the squeezing
phase $\delta\theta$, but for practical squeezing parameters in current
experiments, usually $\sinh2r\ll\cosh2r$, thus still the first increasing
term $\mathsf{X}_{t}$ dominates, and $\dot{S}_{\text{\textsc{b}}}$
keeps positive.

In the strong squeezing limit ($r\gg1$), $\dot{S}_{\text{\textsc{b}}}\lesssim\frac{\lambda\Omega_{c}^{2}}{\pi T}\cosh2r\cdot f(t,\,\delta\theta)$,
where $f(t,\,\delta\theta):=\mathsf{X}_{t}-(\mathsf{Y}_{t}\cos\delta\theta+\mathsf{Z}_{t}\sin\delta\theta)$
{[}see Fig.\,\ref{fig-squ}(c, d){]}. In most area, $f(t,\,\delta\theta)$
keeps positive, but for certain phase $\delta\theta$ {[}the shadowed
area in Fig.\,\ref{fig-squ}(c){]}, $f(t,\,\delta\theta)$ could
become a small but negative value, indicating the decreasing of the
bath entropy in this area. However, we also should notice that the
time scale in Fig.\,\ref{fig-squ}(c, d) is around $t\sim\Omega_{c}^{-1}$,
which is a very shot comparing with the system time scale ($\sim\kappa^{-1}$),
and this is just the relaxation time of the bath. In the coarse-grained
time scale (Markovian approximation), this decreasing of bath entropy
is negligible.

For zero temperature case, the total S-B system always stays in a
pure state. With the time evolves, $\rho_{\text{\textsc{sb}}}(t)$
becomes a pure entangled state, thus we always have $\dot{S}_{\text{\textsc{b}}}(t)=\dot{S}_{\text{\textsc{s}}}(t)$,
which also increases monotonically, as already discussed in Sec.\,III.
In this case, the thermal fluctuation does not contribute to the bath
entropy, and $S_{\text{\textsc{b}}}$ all comes from the correlating
with the TLS.

\section{Summary}

In this paper, we study the entropy dynamics of a dephasing model,
where a TLS is coupled with a squeezed thermal bath via non-demolition
interaction. We show the exact evolution operator, and the time dependent
states of both the TLS and its bath can be obtained exactly. Based
on these states, we calculate the entropy dynamics of both the TLS
and the bath, and find that the dephasing rate of the system relies
on the squeezing phase of the bath. In zero temperature and high temperature
limits, both the system and bath entropy increases monotonically in
the coarse-grained time scale (Markovian approximation).

Moreover, we find that the dephasing rate of the system relies on
the squeezing phase $\theta$ of the bath. Particularly, in the high
temperature limit, the system dephasing process is Markovian, and
the dephasing rate $\kappa=2\lambda T[\cosh2r-\frac{\ln4}{\pi}\sinh2r\sin\delta\theta]$.
We also notice that this dephasing rate cannot be precisely given
by the Born-Markovian approximation which is widely adopted in open
quantum systems.

We also discuss the validity of using the thermal entropy analogy
$\dot{S}_{\text{\textsc{b}}}\approx\frac{d}{dt}\langle\hat{H}_{\text{\textsc{b}}}\rangle/T$
to approximately calculate the bath entropy. For this dephasing model,
when the bath temperature is high, the thermal fluctuation dominates
the bath entropy dynamics, and this approximation works well; in the
low temperature regime, the non-Gaussian property of the bath state
becomes more important, and this approximation does not work well.

\emph{Acknowledgement} - S.-W. Li appreciates a lot for the helpful
discussion with G. S. Agarwal about the exact solution of the open
quantum system. This study is supported by the Office of Naval Research
(Award No. N00014-16-1-3054) and the Robert A. Welch Foundation (Grant
No. A-1261).

\appendix

\section{Evolution operator \label{sec:Evolution-operator}}

Here we show the derivation of the evolution operator $U_{I}(t)$
{[}Eq.\,(\ref{eq:U-evo}){]} \cite{schaller_open_2014}. In the interaction
picture, the evolution operator can be written as the following time-ordered
form 
\begin{align}
U_{I}(t) & =\mathbf{T}\exp[-i\int_{0}^{t}ds\,\hat{V}_{\text{\textsc{sb}}}(s)]\nonumber \\
 & =\mathbf{T}\lim_{N\rightarrow\infty}\exp[-i\sum_{n=0}^{N-1}\hat{V}_{\text{\textsc{sb}}}(t_{n})\delta t],
\end{align}
where $\delta t=t/N$ is the time interval, $t_{n}=n\delta t$, and
$\hat{V}_{\text{\textsc{sb}}}(t)$ is in the interaction picture.

For the interaction Hamiltonian (\ref{eq:V_SB}) in the above dephasing
model, we have $[\hat{V}_{\text{\textsc{sb}}}(t),\,\hat{V}_{\text{\textsc{sb}}}(s)]=2i\sum_{k}|g_{k}|^{2}\sin\omega_{k}(s-t)$.
This is a c-number, thus we can use the Baker-Campbell-Hausdorff (BCH)
formula:
\begin{equation}
e^{\sum_{n=1}^{N}A_{n}}=e^{A_{1}}e^{A_{2}}\cdots e^{A_{N}}e^{-\frac{1}{2}\sum_{m<n}[A_{m},A_{n}]}.
\end{equation}
Then we obtain
\[
U_{I}(t)=\mathbf{T}\lim_{N\rightarrow\infty}\prod_{n=0}^{N-1}e^{-i\hat{V}_{\text{\textsc{sb}}}(t_{n})\delta t}\cdot e^{\frac{\delta t^{2}}{2}\sum_{m<n}[\hat{V}_{\text{\textsc{sb}}}(t_{m}),\hat{V}_{\text{\textsc{sb}}}(t_{n})]}.
\]
The above product is already time-ordered, thus the time-order operator
$\mathbf{T}$ can be removed. Then using the BCH formula reversely,
the evolution operator becomes
\begin{align}
U_{I}(t) & =\lim_{N\rightarrow\infty}\exp[-i\sum_{n=0}^{N-1}\hat{V}_{\text{\textsc{sb}}}(t_{n})\delta t]=\exp[-i\int_{0}^{t}ds\,\hat{V}_{\text{\textsc{sb}}}(s)]\nonumber \\
 & =\exp\big\{\hat{\sigma}^{z}\cdot[\alpha_{k}(t)b_{k}^{\dagger}-\alpha_{k}^{\ast}(t)b_{k}]\big\},
\end{align}
where $\alpha_{k}(t)=\mu_{k}(1-e^{i\omega_{k}t})$ and $\mu_{k}=g_{k}/\omega_{k}$.

\section{Markovian master equation \label{sec:Markovian}}

Here we use the Born-Markovian approximation to derive a master equation
for the TLS \cite{breuer_theory_2002,li_steady_2015}. The master
equation is derived from 
\begin{equation}
\dot{\rho}_{\text{\textsc{s}}}=-\mathrm{tr}_{\text{\textsc{b}}}\int_{0}^{\infty}ds\,[\hat{V}_{\text{\textsc{sb}}}(t),[\hat{V}_{\text{\textsc{sb}}}(t-s),\rho_{\text{\textsc{s}}}(t)\otimes\rho_{\text{\textsc{b}}}(0)]]
\end{equation}
where $\hat{V}_{\text{\textsc{sb}}}(t)=\hat{\sigma}^{z}\cdot({\displaystyle \sum_{k}}g_{k}\hat{b}_{k}e^{-i\omega_{k}t}+g_{k}^{*}\hat{b}_{k}^{\dagger}e^{i\omega_{k}t}):=\hat{\sigma}^{z}\cdot\hat{B}(t)$.
The above commutator gives
\begin{align*}
 & \int_{0}^{\infty}ds\,\big\langle\hat{B}^{\dagger}(t)\hat{B}(t-s)\big\rangle[\hat{\sigma}^{z}\rho_{\text{\textsc{s}}}(t),\,\hat{\sigma}^{z}]\\
= & \sum_{k}\int_{0}^{\infty}ds\,\big\langle(g_{k}^{*}\hat{b}_{k}^{\dagger}e^{i\omega_{k}t}+g_{k}\hat{b}_{k}e^{-i\omega_{k}t})\\
 & \qquad\times(g_{k}\hat{b}_{k}e^{-i\omega_{k}(t-s)}+g_{k}^{*}\hat{b}_{k}^{\dagger}e^{i\omega_{k}(t-s)})\big\rangle\cdot[\hat{\sigma}^{z}\rho,\hat{\sigma}^{z}]\\
= & \int_{0}^{\infty}ds\int_{0}^{\infty}\frac{d\omega}{2\pi}\,J(\omega)\Big([\tilde{\mathsf{n}}(\omega)e^{i\omega s}+(\tilde{\mathsf{n}}(\omega)+1)e^{-i\omega s}]\\
 & \qquad+[\tilde{\mathsf{u}}(\omega)e^{-2i\omega t+i\omega s}+\mathbf{h.c.}]\Big)[\sigma_{z}\rho,\sigma_{z}],
\end{align*}
where we denote
\begin{align}
\tilde{\mathsf{n}}(\omega) & =\cosh2r[\overline{\mathsf{n}}_{\mathrm{p}}(\omega)+\frac{1}{2}]-\frac{1}{2},\nonumber \\
\tilde{\mathsf{u}}(\omega) & =-e^{i\delta\theta}\sinh2r[\overline{\mathsf{n}}_{\mathrm{p}}(\omega)+\frac{1}{2}],
\end{align}
 and $\overline{\mathsf{n}}_{\mathrm{p}}(\omega)=[\exp(\omega/T)-1]^{-1}$
is the Planck function.

Utilizing the formula
\begin{equation}
\int_{0}^{\infty}ds\,e^{i(\varepsilon-\omega)s}=\pi\delta(\varepsilon-\omega)+i\mathbf{P}\frac{1}{\varepsilon-\omega},
\end{equation}
we obtain (omitting the principle integral)
\begin{align}
 & \int_{0}^{\infty}ds\,\big\langle\hat{B}^{\dagger}(t)\hat{B}(t-s)\big\rangle\nonumber \\
= & \lim_{\omega\rightarrow0}J(\omega)\Big([\tilde{\mathsf{n}}(\omega)+\frac{1}{2}]+\frac{1}{2}[\tilde{\mathsf{u}}(\omega)e^{-2i\omega t}+\mathbf{h.c.}]\Big).
\end{align}
Adopting the Ohmic spectrum $J(\omega)=\lambda\omega\,e^{-\frac{\omega}{\Omega_{c}}}$,
we have 
\begin{equation}
\lim_{\omega\rightarrow0}\lambda\omega\,e^{-\frac{\omega}{\Omega_{c}}}[\overline{\mathsf{n}}_{\mathrm{p}}(\omega)+\frac{1}{2}]=\lambda T.
\end{equation}
 Now we obtain the master equation $\dot{\rho}_{\text{\textsc{s}}}={\cal L}[\rho_{\text{\textsc{s}}}]$
where
\begin{align}
{\cal L}[\rho_{\text{\textsc{s}}}] & =\frac{1}{2}\kappa'\Big([\sigma_{z}\rho_{\text{\textsc{s}}},\,\sigma_{z}]+[\sigma_{z},\,\rho_{\text{\textsc{s}}}\sigma_{z}]\Big),\nonumber \\
\kappa' & =2\lambda T(\cosh2r-\sinh2r\cos\delta\theta).
\end{align}
Notice that here the phase dependence in the dephasing rate $\kappa'$
is different from what we obtained in the main text using the exact
evolution operator, which is more precise.

\section{The monotonic increase of the decay rate \label{sec:The-increasing-rate}}

Here we show the proof for the monotonic increase of the system decay
factor $\Gamma(t)$ in Sec.\,\ref{subsec:Zero-temperature-case},
namely, $\dot{\Gamma}(t)$ is always positive. Since for any $r>0$,
we have $\cosh2r>\sinh2r>0$, thus 
\begin{align}
 & \dot{\Gamma}(t)\cdot\frac{\pi}{\lambda\cosh2r}\ge\dot{\mathsf{A}}_{t}-\sqrt{\dot{\mathsf{B}}_{t}^{2}+\dot{\mathsf{C}}_{t}^{2}}\Big(\frac{\dot{\mathsf{B}}_{t}}{\sqrt{\dot{\mathsf{B}}_{t}^{2}+\dot{\mathsf{C}}_{t}^{2}}}\cos\delta\theta\nonumber \\
+ & \frac{\dot{\mathsf{C}}_{t}}{\sqrt{\dot{\mathsf{B}}_{t}^{2}+\dot{\mathsf{C}}_{t}^{2}}}\sin\delta\theta\Big)=\dot{\mathsf{A}}_{t}-\sqrt{\dot{\mathsf{B}}_{t}^{2}+\dot{\mathsf{C}}_{t}^{2}}\cos(\delta\theta-\varphi)\nonumber \\
\ge & \dot{\mathsf{A}}_{t}-\sqrt{\dot{\mathsf{B}}_{t}^{2}+\dot{\mathsf{C}}_{t}^{2}},
\end{align}
where we denote $\varphi:=\tan^{\text{-1}}(\dot{\mathsf{C}}_{t}/\dot{\mathsf{B}}_{t})$
and $\tau:=\Omega_{c}t$. Using the expression of $\mathsf{A}_{t}$,
$\mathsf{B}_{t}$ and $\mathsf{C}_{t}$ {[}Eq.\,(\ref{eq:ABC-0}){]},
we have 
\begin{equation}
\dot{\mathsf{A}}_{t}-\sqrt{\dot{\mathsf{B}}_{t}^{2}+\dot{\mathsf{C}}_{t}^{2}}=\frac{2\tau}{1+\tau^{2}}\Big[1-\sqrt{\frac{1+\tau^{2}}{1+4\tau^{2}}}\Big]\ge0.
\end{equation}
Thus, in the area $\tau\ge0$, and we always have $\dot{\Gamma}(t)\ge0$.
Therefore, $\Gamma(t)$ increases monotonically, and is always positive
(since $\Gamma(0)=0$).

\bibliographystyle{apsrev4-1}
\bibliography{Refs}

\begin{thebibliography}{47}%
\makeatletter
\providecommand \@ifxundefined [1]{%
 \@ifx{#1\undefined}
}%
\providecommand \@ifnum [1]{%
 \ifnum #1\expandafter \@firstoftwo
 \else \expandafter \@secondoftwo
 \fi
}%
\providecommand \@ifx [1]{%
 \ifx #1\expandafter \@firstoftwo
 \else \expandafter \@secondoftwo
 \fi
}%
\providecommand \natexlab [1]{#1}%
\providecommand \enquote  [1]{``#1''}%
\providecommand \bibnamefont  [1]{#1}%
\providecommand \bibfnamefont [1]{#1}%
\providecommand \citenamefont [1]{#1}%
\providecommand \href@noop [0]{\@secondoftwo}%
\providecommand \href [0]{\begingroup \@sanitize@url \@href}%
\providecommand \@href[1]{\@@startlink{#1}\@@href}%
\providecommand \@@href[1]{\endgroup#1\@@endlink}%
\providecommand \@sanitize@url [0]{\catcode `\\12\catcode `\$12\catcode
  `\&12\catcode `\#12\catcode `\^12\catcode `\_12\catcode `\%12\relax}%
\providecommand \@@startlink[1]{}%
\providecommand \@@endlink[0]{}%
\providecommand \url  [0]{\begingroup\@sanitize@url \@url }%
\providecommand \@url [1]{\endgroup\@href {#1}{\urlprefix }}%
\providecommand \urlprefix  [0]{URL }%
\providecommand \Eprint [0]{\href }%
\providecommand \doibase [0]{http://dx.doi.org/}%
\providecommand \selectlanguage [0]{\@gobble}%
\providecommand \bibinfo  [0]{\@secondoftwo}%
\providecommand \bibfield  [0]{\@secondoftwo}%
\providecommand \translation [1]{[#1]}%
\providecommand \BibitemOpen [0]{}%
\providecommand \bibitemStop [0]{}%
\providecommand \bibitemNoStop [0]{.\EOS\space}%
\providecommand \EOS [0]{\spacefactor3000\relax}%
\providecommand \BibitemShut  [1]{\csname bibitem#1\endcsname}%
\let\auto@bib@innerbib\@empty
\bibitem [{\citenamefont {Spohn}(1978)}]{spohn_entropy_1978}%
  \BibitemOpen
  \bibfield  {author} {\bibinfo {author} {\bibfnamefont {H.}~\bibnamefont
  {Spohn}},\ }\href {\doibase 10.1063/1.523789} {\bibfield  {journal} {\bibinfo
   {journal} {J. Math. Phys.}\ }\textbf {\bibinfo {volume} {19}},\ \bibinfo
  {pages} {1227} (\bibinfo {year} {1978})}\BibitemShut {NoStop}%
\bibitem [{\citenamefont {Spohn}\ and\ \citenamefont
  {Lebowitz}(1978)}]{spohn_irreversible_1978}%
  \BibitemOpen
  \bibfield  {author} {\bibinfo {author} {\bibfnamefont {H.}~\bibnamefont
  {Spohn}}\ and\ \bibinfo {author} {\bibfnamefont {J.~L.}\ \bibnamefont
  {Lebowitz}},\ }in\ \href
  {http://onlinelibrary.wiley.com/doi/10.1002/9780470142578.ch2/summary} {\emph
  {\bibinfo {booktitle} {Advances in {Chemical} {Physics}}}},\ \bibinfo
  {editor} {edited by\ \bibinfo {editor} {\bibfnamefont {S.~A.}\ \bibnamefont
  {Rice}}}\ (\bibinfo  {publisher} {John Wiley \& Sons, Inc.},\ \bibinfo {year}
  {1978})\ pp.\ \bibinfo {pages} {109--142}\BibitemShut {NoStop}%
\bibitem [{\citenamefont {Quan}\ \emph {et~al.}(2005)\citenamefont {Quan},
  \citenamefont {Zhang},\ and\ \citenamefont {Sun}}]{quan_quantum_2005}%
  \BibitemOpen
  \bibfield  {author} {\bibinfo {author} {\bibfnamefont {H.~T.}\ \bibnamefont
  {Quan}}, \bibinfo {author} {\bibfnamefont {P.}~\bibnamefont {Zhang}}, \ and\
  \bibinfo {author} {\bibfnamefont {C.~P.}\ \bibnamefont {Sun}},\ }\href
  {\doibase 10.1103/PhysRevE.72.056110} {\bibfield  {journal} {\bibinfo
  {journal} {Phys. Rev. E}\ }\textbf {\bibinfo {volume} {72}},\ \bibinfo
  {pages} {056110} (\bibinfo {year} {2005})}\BibitemShut {NoStop}%
\bibitem [{\citenamefont {Quan}\ \emph {et~al.}(2006)\citenamefont {Quan},
  \citenamefont {Zhang},\ and\ \citenamefont
  {Sun}}]{quan_quantum-classical_2006}%
  \BibitemOpen
  \bibfield  {author} {\bibinfo {author} {\bibfnamefont {H.~T.}\ \bibnamefont
  {Quan}}, \bibinfo {author} {\bibfnamefont {P.}~\bibnamefont {Zhang}}, \ and\
  \bibinfo {author} {\bibfnamefont {C.~P.}\ \bibnamefont {Sun}},\ }\href
  {\doibase 10.1103/PhysRevE.73.036122} {\bibfield  {journal} {\bibinfo
  {journal} {Phys. Rev. E}\ }\textbf {\bibinfo {volume} {73}},\ \bibinfo
  {pages} {036122} (\bibinfo {year} {2006})}\BibitemShut {NoStop}%
\bibitem [{\citenamefont {Scully}\ \emph {et~al.}(2003)\citenamefont {Scully},
  \citenamefont {Zubairy}, \citenamefont {Agarwal},\ and\ \citenamefont
  {Walther}}]{scully_extracting_2003}%
  \BibitemOpen
  \bibfield  {author} {\bibinfo {author} {\bibfnamefont {M.~O.}\ \bibnamefont
  {Scully}}, \bibinfo {author} {\bibfnamefont {M.~S.}\ \bibnamefont {Zubairy}},
  \bibinfo {author} {\bibfnamefont {G.~S.}\ \bibnamefont {Agarwal}}, \ and\
  \bibinfo {author} {\bibfnamefont {H.}~\bibnamefont {Walther}},\ }\href
  {\doibase 10.1126/science.1078955} {\bibfield  {journal} {\bibinfo  {journal}
  {Science}\ }\textbf {\bibinfo {volume} {299}},\ \bibinfo {pages} {862}
  (\bibinfo {year} {2003})}\BibitemShut {NoStop}%
\bibitem [{\citenamefont {Ro{\ss}nagel}\ \emph {et~al.}(2014)\citenamefont
  {Ro{\ss}nagel}, \citenamefont {Abah}, \citenamefont {Schmidt-Kaler},
  \citenamefont {Singer},\ and\ \citenamefont
  {Lutz}}]{rosnagel_nanoscale_2014}%
  \BibitemOpen
  \bibfield  {author} {\bibinfo {author} {\bibfnamefont {J.}~\bibnamefont
  {Ro{\ss}nagel}}, \bibinfo {author} {\bibfnamefont {O.}~\bibnamefont {Abah}},
  \bibinfo {author} {\bibfnamefont {F.}~\bibnamefont {Schmidt-Kaler}}, \bibinfo
  {author} {\bibfnamefont {K.}~\bibnamefont {Singer}}, \ and\ \bibinfo {author}
  {\bibfnamefont {E.}~\bibnamefont {Lutz}},\ }\href {\doibase
  10.1103/PhysRevLett.112.030602} {\bibfield  {journal} {\bibinfo  {journal}
  {Phys. Rev. Lett.}\ }\textbf {\bibinfo {volume} {112}},\ \bibinfo {pages}
  {030602} (\bibinfo {year} {2014})}\BibitemShut {NoStop}%
\bibitem [{\citenamefont {Correa}\ \emph {et~al.}(2014)\citenamefont {Correa},
  \citenamefont {Palao}, \citenamefont {Alonso},\ and\ \citenamefont
  {Adesso}}]{correa_quantum-enhanced_2014}%
  \BibitemOpen
  \bibfield  {author} {\bibinfo {author} {\bibfnamefont {L.~A.}\ \bibnamefont
  {Correa}}, \bibinfo {author} {\bibfnamefont {J.~P.}\ \bibnamefont {Palao}},
  \bibinfo {author} {\bibfnamefont {D.}~\bibnamefont {Alonso}}, \ and\ \bibinfo
  {author} {\bibfnamefont {G.}~\bibnamefont {Adesso}},\ }\href
  {http://dx.doi.org/10.1038/srep03949} {\bibfield  {journal} {\bibinfo
  {journal} {Sci. Rep.}\ }\textbf {\bibinfo {volume} {4}},\ \bibinfo {pages}
  {3949} (\bibinfo {year} {2014})}\BibitemShut {NoStop}%
\bibitem [{\citenamefont {Long}\ and\ \citenamefont
  {Liu}(2015)}]{long_performance_2015}%
  \BibitemOpen
  \bibfield  {author} {\bibinfo {author} {\bibfnamefont {R.}~\bibnamefont
  {Long}}\ and\ \bibinfo {author} {\bibfnamefont {W.}~\bibnamefont {Liu}},\
  }\href {\doibase 10.1103/PhysRevE.91.062137} {\bibfield  {journal} {\bibinfo
  {journal} {Phys. Rev. E}\ }\textbf {\bibinfo {volume} {91}},\ \bibinfo
  {pages} {062137} (\bibinfo {year} {2015})}\BibitemShut {NoStop}%
\bibitem [{\citenamefont {Agarwalla}\ \emph {et~al.}(2017)\citenamefont
  {Agarwalla}, \citenamefont {Jiang},\ and\ \citenamefont
  {Segal}}]{agarwalla_quantum_2017}%
  \BibitemOpen
  \bibfield  {author} {\bibinfo {author} {\bibfnamefont {B.~K.}\ \bibnamefont
  {Agarwalla}}, \bibinfo {author} {\bibfnamefont {J.-H.}\ \bibnamefont
  {Jiang}}, \ and\ \bibinfo {author} {\bibfnamefont {D.}~\bibnamefont
  {Segal}},\ }\href {\doibase 10.1103/PhysRevB.96.104304} {\bibfield  {journal}
  {\bibinfo  {journal} {Phys. Rev. B}\ }\textbf {\bibinfo {volume} {96}},\
  \bibinfo {pages} {104304} (\bibinfo {year} {2017})}\BibitemShut {NoStop}%
\bibitem [{\citenamefont {Gelbwaser-Klimovsky}\ and\ \citenamefont
  {Kurizki}(2014)}]{gelbwaser-klimovsky_heat-machine_2014}%
  \BibitemOpen
  \bibfield  {author} {\bibinfo {author} {\bibfnamefont {D.}~\bibnamefont
  {Gelbwaser-Klimovsky}}\ and\ \bibinfo {author} {\bibfnamefont
  {G.}~\bibnamefont {Kurizki}},\ }\href {\doibase 10.1103/PhysRevE.90.022102}
  {\bibfield  {journal} {\bibinfo  {journal} {Phys. Rev. E}\ }\textbf {\bibinfo
  {volume} {90}},\ \bibinfo {pages} {022102} (\bibinfo {year}
  {2014})}\BibitemShut {NoStop}%
\bibitem [{\citenamefont {Gardas}\ and\ \citenamefont
  {Deffner}(2015)}]{gardas_thermodynamic_2015}%
  \BibitemOpen
  \bibfield  {author} {\bibinfo {author} {\bibfnamefont {B.}~\bibnamefont
  {Gardas}}\ and\ \bibinfo {author} {\bibfnamefont {S.}~\bibnamefont
  {Deffner}},\ }\href {\doibase 10.1103/PhysRevE.92.042126} {\bibfield
  {journal} {\bibinfo  {journal} {Phys. Rev. E}\ }\textbf {\bibinfo {volume}
  {92}},\ \bibinfo {pages} {042126} (\bibinfo {year} {2015})}\BibitemShut
  {NoStop}%
\bibitem [{\citenamefont {Da{\u g}}\ \emph {et~al.}(2016)\citenamefont {Da{\u
  g}}, \citenamefont {Niedenzu}, \citenamefont {M{\"u}stecapl{\i}o{\u g}lu},\
  and\ \citenamefont {Kurizki}}]{dag_multiatom_2016}%
  \BibitemOpen
  \bibfield  {author} {\bibinfo {author} {\bibfnamefont {C.~B.}\ \bibnamefont
  {Da{\u g}}}, \bibinfo {author} {\bibfnamefont {W.}~\bibnamefont {Niedenzu}},
  \bibinfo {author} {\bibfnamefont {{\"O}.~E.}\ \bibnamefont
  {M{\"u}stecapl{\i}o{\u g}lu}}, \ and\ \bibinfo {author} {\bibfnamefont
  {G.}~\bibnamefont {Kurizki}},\ }\href {\doibase 10.3390/e18070244} {\bibfield
   {journal} {\bibinfo  {journal} {Entropy}\ }\textbf {\bibinfo {volume}
  {18}},\ \bibinfo {pages} {244} (\bibinfo {year} {2016})}\BibitemShut
  {NoStop}%
\bibitem [{\citenamefont {Zhang}(2008)}]{zhang_general_2008}%
  \BibitemOpen
  \bibfield  {author} {\bibinfo {author} {\bibfnamefont {Q.-R.}\ \bibnamefont
  {Zhang}},\ }\href {\doibase 10.1142/S0218301308009859} {\bibfield  {journal}
  {\bibinfo  {journal} {Int. J. Mod. Phys. E}\ }\textbf {\bibinfo {volume}
  {17}},\ \bibinfo {pages} {531} (\bibinfo {year} {2008})},\ \bibinfo {note}
  {arXiv:quant-ph/0610005}\BibitemShut {NoStop}%
\bibitem [{\citenamefont {Zhang}(2009)}]{zhang_information_2009}%
  \BibitemOpen
  \bibfield  {author} {\bibinfo {author} {\bibfnamefont {Q.-R.}\ \bibnamefont
  {Zhang}},\ }\href {\doibase 10.1016/j.physa.2009.06.039} {\bibfield
  {journal} {\bibinfo  {journal} {Physica A}\ }\textbf {\bibinfo {volume}
  {388}},\ \bibinfo {pages} {4041} (\bibinfo {year} {2009})}\BibitemShut
  {NoStop}%
\bibitem [{\citenamefont {Hilt}\ and\ \citenamefont
  {Lutz}(2009)}]{hilt_system-bath_2009}%
  \BibitemOpen
  \bibfield  {author} {\bibinfo {author} {\bibfnamefont {S.}~\bibnamefont
  {Hilt}}\ and\ \bibinfo {author} {\bibfnamefont {E.}~\bibnamefont {Lutz}},\
  }\href {\doibase 10.1103/PhysRevA.79.010101} {\bibfield  {journal} {\bibinfo
  {journal} {Phys. Rev. A}\ }\textbf {\bibinfo {volume} {79}},\ \bibinfo
  {pages} {010101} (\bibinfo {year} {2009})}\BibitemShut {NoStop}%
\bibitem [{\citenamefont {Esposito}\ \emph {et~al.}(2010)\citenamefont
  {Esposito}, \citenamefont {Lindenberg},\ and\ \citenamefont
  {Broeck}}]{esposito_entropy_2010}%
  \BibitemOpen
  \bibfield  {author} {\bibinfo {author} {\bibfnamefont {M.}~\bibnamefont
  {Esposito}}, \bibinfo {author} {\bibfnamefont {K.}~\bibnamefont
  {Lindenberg}}, \ and\ \bibinfo {author} {\bibfnamefont {C.~V.~d.}\
  \bibnamefont {Broeck}},\ }\href {\doibase 10.1088/1367-2630/12/1/013013}
  {\bibfield  {journal} {\bibinfo  {journal} {New J. Phys.}\ }\textbf {\bibinfo
  {volume} {12}},\ \bibinfo {pages} {013013} (\bibinfo {year}
  {2010})}\BibitemShut {NoStop}%
\bibitem [{\citenamefont {Pernice}\ and\ \citenamefont
  {Strunz}(2011)}]{pernice_decoherence_2011}%
  \BibitemOpen
  \bibfield  {author} {\bibinfo {author} {\bibfnamefont {A.}~\bibnamefont
  {Pernice}}\ and\ \bibinfo {author} {\bibfnamefont {W.~T.}\ \bibnamefont
  {Strunz}},\ }\href {\doibase 10.1103/PhysRevA.84.062121} {\bibfield
  {journal} {\bibinfo  {journal} {Phys. Rev. A}\ }\textbf {\bibinfo {volume}
  {84}},\ \bibinfo {pages} {062121} (\bibinfo {year} {2011})}\BibitemShut
  {NoStop}%
\bibitem [{\citenamefont {Pucci}\ \emph {et~al.}(2013)\citenamefont {Pucci},
  \citenamefont {Esposito},\ and\ \citenamefont {Peliti}}]{pucci_entropy_2013}%
  \BibitemOpen
  \bibfield  {author} {\bibinfo {author} {\bibfnamefont {L.}~\bibnamefont
  {Pucci}}, \bibinfo {author} {\bibfnamefont {M.}~\bibnamefont {Esposito}}, \
  and\ \bibinfo {author} {\bibfnamefont {L.}~\bibnamefont {Peliti}},\ }\href
  {\doibase 10.1088/1742-5468/2013/04/P04005} {\bibfield  {journal} {\bibinfo
  {journal} {J. Stat. Mech.}\ }\textbf {\bibinfo {volume} {2013}},\ \bibinfo
  {pages} {P04005} (\bibinfo {year} {2013})}\BibitemShut {NoStop}%
\bibitem [{\citenamefont {Manzano}\ \emph {et~al.}(2016)\citenamefont
  {Manzano}, \citenamefont {Galve}, \citenamefont {Zambrini},\ and\
  \citenamefont {Parrondo}}]{manzano_entropy_2016}%
  \BibitemOpen
  \bibfield  {author} {\bibinfo {author} {\bibfnamefont {G.}~\bibnamefont
  {Manzano}}, \bibinfo {author} {\bibfnamefont {F.}~\bibnamefont {Galve}},
  \bibinfo {author} {\bibfnamefont {R.}~\bibnamefont {Zambrini}}, \ and\
  \bibinfo {author} {\bibfnamefont {J.~M.~R.}\ \bibnamefont {Parrondo}},\
  }\href {\doibase 10.1103/PhysRevE.93.052120} {\bibfield  {journal} {\bibinfo
  {journal} {Phys. Rev. E}\ }\textbf {\bibinfo {volume} {93}},\ \bibinfo
  {pages} {052120} (\bibinfo {year} {2016})}\BibitemShut {NoStop}%
\bibitem [{\citenamefont {Alipour}\ \emph {et~al.}(2016)\citenamefont
  {Alipour}, \citenamefont {Benatti}, \citenamefont {Bakhshinezhad},
  \citenamefont {Afsary}, \citenamefont {Marcantoni},\ and\ \citenamefont
  {Rezakhani}}]{alipour_correlations_2016}%
  \BibitemOpen
  \bibfield  {author} {\bibinfo {author} {\bibfnamefont {S.}~\bibnamefont
  {Alipour}}, \bibinfo {author} {\bibfnamefont {F.}~\bibnamefont {Benatti}},
  \bibinfo {author} {\bibfnamefont {F.}~\bibnamefont {Bakhshinezhad}}, \bibinfo
  {author} {\bibfnamefont {M.}~\bibnamefont {Afsary}}, \bibinfo {author}
  {\bibfnamefont {S.}~\bibnamefont {Marcantoni}}, \ and\ \bibinfo {author}
  {\bibfnamefont {A.~T.}\ \bibnamefont {Rezakhani}},\ }\href {\doibase
  10.1038/srep35568} {\bibfield  {journal} {\bibinfo  {journal} {Sci. Rep.}\
  }\textbf {\bibinfo {volume} {6}},\ \bibinfo {pages} {35568} (\bibinfo {year}
  {2016})}\BibitemShut {NoStop}%
\bibitem [{\citenamefont {Li}(2017)}]{li_production_2017}%
  \BibitemOpen
  \bibfield  {author} {\bibinfo {author} {\bibfnamefont {S.-W.}\ \bibnamefont
  {Li}},\ }\href {\doibase 10.1103/PhysRevE.96.012139} {\bibfield  {journal}
  {\bibinfo  {journal} {Phys. Rev. E}\ }\textbf {\bibinfo {volume} {96}},\
  \bibinfo {pages} {012139} (\bibinfo {year} {2017})},\ \bibinfo {note} {arXiv:
  1612.03884}\BibitemShut {NoStop}%
\bibitem [{\citenamefont {Strasberg}\ \emph {et~al.}(2017)\citenamefont
  {Strasberg}, \citenamefont {Schaller}, \citenamefont {Brandes},\ and\
  \citenamefont {Esposito}}]{strasberg_quantum_2017}%
  \BibitemOpen
  \bibfield  {author} {\bibinfo {author} {\bibfnamefont {P.}~\bibnamefont
  {Strasberg}}, \bibinfo {author} {\bibfnamefont {G.}~\bibnamefont {Schaller}},
  \bibinfo {author} {\bibfnamefont {T.}~\bibnamefont {Brandes}}, \ and\
  \bibinfo {author} {\bibfnamefont {M.}~\bibnamefont {Esposito}},\ }\href
  {\doibase 10.1103/PhysRevX.7.021003} {\bibfield  {journal} {\bibinfo
  {journal} {Phys. Rev. X}\ }\textbf {\bibinfo {volume} {7}},\ \bibinfo {pages}
  {021003} (\bibinfo {year} {2017})}\BibitemShut {NoStop}%
\bibitem [{\citenamefont {Iyoda}\ \emph {et~al.}(2017)\citenamefont {Iyoda},
  \citenamefont {Kaneko},\ and\ \citenamefont
  {Sagawa}}]{iyoda_fluctuation_2017}%
  \BibitemOpen
  \bibfield  {author} {\bibinfo {author} {\bibfnamefont {E.}~\bibnamefont
  {Iyoda}}, \bibinfo {author} {\bibfnamefont {K.}~\bibnamefont {Kaneko}}, \
  and\ \bibinfo {author} {\bibfnamefont {T.}~\bibnamefont {Sagawa}},\ }\href
  {\doibase 10.1103/PhysRevLett.119.100601} {\bibfield  {journal} {\bibinfo
  {journal} {Phys. Rev. Lett.}\ }\textbf {\bibinfo {volume} {119}},\ \bibinfo
  {pages} {100601} (\bibinfo {year} {2017})}\BibitemShut {NoStop}%
\bibitem [{\citenamefont {Scully}\ \emph {et~al.}(2017)\citenamefont {Scully},
  \citenamefont {Fulling}, \citenamefont {Lee}, \citenamefont {Page},
  \citenamefont {Schleich},\ and\ \citenamefont
  {Svidzinsky}}]{scully_radiation_2017}%
  \BibitemOpen
  \bibfield  {author} {\bibinfo {author} {\bibfnamefont {M.~O.}\ \bibnamefont
  {Scully}}, \bibinfo {author} {\bibfnamefont {S.}~\bibnamefont {Fulling}},
  \bibinfo {author} {\bibfnamefont {D.}~\bibnamefont {Lee}}, \bibinfo {author}
  {\bibfnamefont {D.}~\bibnamefont {Page}}, \bibinfo {author} {\bibfnamefont
  {W.}~\bibnamefont {Schleich}}, \ and\ \bibinfo {author} {\bibfnamefont
  {A.}~\bibnamefont {Svidzinsky}},\ }\href {http://arxiv.org/abs/1709.00481}
  {\bibfield  {journal} {\bibinfo  {journal} {arXiv:1709.00481}\ } (\bibinfo
  {year} {2017})}\BibitemShut {NoStop}%
\bibitem [{\citenamefont {Dong}\ \emph {et~al.}(2016)\citenamefont {Dong},
  \citenamefont {Li}, \citenamefont {Yi}, \citenamefont {Agarwal},\ and\
  \citenamefont {Scully}}]{dong_photon-blockade_2016}%
  \BibitemOpen
  \bibfield  {author} {\bibinfo {author} {\bibfnamefont {H.}~\bibnamefont
  {Dong}}, \bibinfo {author} {\bibfnamefont {S.-W.}\ \bibnamefont {Li}},
  \bibinfo {author} {\bibfnamefont {Z.}~\bibnamefont {Yi}}, \bibinfo {author}
  {\bibfnamefont {G.~S.}\ \bibnamefont {Agarwal}}, \ and\ \bibinfo {author}
  {\bibfnamefont {M.~O.}\ \bibnamefont {Scully}},\ }\href
  {http://arxiv.org/abs/1608.04364} {\bibfield  {journal} {\bibinfo  {journal}
  {arXiv:1608.04364}\ } (\bibinfo {year} {2016})}\BibitemShut {NoStop}%
\bibitem [{\citenamefont {May}\ and\ \citenamefont
  {K{\"u}hn}(2000)}]{may_charge_2000}%
  \BibitemOpen
  \bibfield  {author} {\bibinfo {author} {\bibfnamefont {V.}~\bibnamefont
  {May}}\ and\ \bibinfo {author} {\bibfnamefont {O.}~\bibnamefont {K{\"u}hn}},\
  }\href@noop {} {\emph {\bibinfo {title} {Charge and {Energy} {Transfer}
  {Dynamics} in {Molecular} {Systems}: {A} {Theoretical} {Introduction}}}},\
  \bibinfo {edition} {1st}\ ed.\ (\bibinfo  {publisher} {Wiley-VCH},\ \bibinfo
  {address} {Berlin ; New York},\ \bibinfo {year} {2000})\BibitemShut {NoStop}%
\bibitem [{\citenamefont {Dong}\ \emph {et~al.}(2017)\citenamefont {Dong},
  \citenamefont {Wang},\ and\ \citenamefont {Kim}}]{dong_how_2017}%
  \BibitemOpen
  \bibfield  {author} {\bibinfo {author} {\bibfnamefont {H.}~\bibnamefont
  {Dong}}, \bibinfo {author} {\bibfnamefont {D.-W.}\ \bibnamefont {Wang}}, \
  and\ \bibinfo {author} {\bibfnamefont {M.~B.}\ \bibnamefont {Kim}},\ }\href
  {http://arxiv.org/abs/1706.02636} {\bibfield  {journal} {\bibinfo  {journal}
  {arXiv:1706.02636}\ } (\bibinfo {year} {2017})}\BibitemShut {NoStop}%
\bibitem [{\citenamefont {Breuer}\ and\ \citenamefont
  {Petruccione}(2002)}]{breuer_theory_2002}%
  \BibitemOpen
  \bibfield  {author} {\bibinfo {author} {\bibfnamefont {H.}~\bibnamefont
  {Breuer}}\ and\ \bibinfo {author} {\bibfnamefont {F.}~\bibnamefont
  {Petruccione}},\ }\href@noop {} {\emph {\bibinfo {title} {The theory of open
  quantum systems}}}\ (\bibinfo  {publisher} {Oxford University Press},\
  \bibinfo {year} {2002})\BibitemShut {NoStop}%
\bibitem [{\citenamefont {Agarwal}(2012)}]{agarwal_quantum_2012}%
  \BibitemOpen
  \bibfield  {author} {\bibinfo {author} {\bibfnamefont {G.~S.}\ \bibnamefont
  {Agarwal}},\ }\href@noop {} {\emph {\bibinfo {title} {Quantum {Optics}}}},\
  \bibinfo {edition} {1st}\ ed.\ (\bibinfo  {publisher} {Cambridge University
  Press},\ \bibinfo {address} {Cambridge, UK},\ \bibinfo {year}
  {2012})\BibitemShut {NoStop}%
\bibitem [{\citenamefont {Schaller}(2014)}]{schaller_open_2014}%
  \BibitemOpen
  \bibfield  {author} {\bibinfo {author} {\bibfnamefont {G.}~\bibnamefont
  {Schaller}},\ }\href@noop {} {\emph {\bibinfo {title} {Open {Quantum}
  {Systems} {Far} from {Equilibrium}}}},\ \bibinfo {edition} {1st}\ ed.\
  (\bibinfo  {publisher} {Springer},\ \bibinfo {address} {New York},\ \bibinfo
  {year} {2014})\BibitemShut {NoStop}%
\bibitem [{\citenamefont {Banerjee}\ and\ \citenamefont
  {Ghosh}(2007)}]{banerjee_dynamics_2007}%
  \BibitemOpen
  \bibfield  {author} {\bibinfo {author} {\bibfnamefont {S.}~\bibnamefont
  {Banerjee}}\ and\ \bibinfo {author} {\bibfnamefont {R.}~\bibnamefont
  {Ghosh}},\ }\href {\doibase 10.1088/1751-8113/40/45/014} {\bibfield
  {journal} {\bibinfo  {journal} {J. Phys. A: Math. Theor.}\ }\textbf {\bibinfo
  {volume} {40}},\ \bibinfo {pages} {13735} (\bibinfo {year}
  {2007})}\BibitemShut {NoStop}%
\bibitem [{\citenamefont {Xu}\ \emph {et~al.}(2014)\citenamefont {Xu},
  \citenamefont {Li}, \citenamefont {Liu},\ and\ \citenamefont
  {Sun}}]{xu_noncanonical_2014}%
  \BibitemOpen
  \bibfield  {author} {\bibinfo {author} {\bibfnamefont {D.~Z.}\ \bibnamefont
  {Xu}}, \bibinfo {author} {\bibfnamefont {S.-W.}\ \bibnamefont {Li}}, \bibinfo
  {author} {\bibfnamefont {X.~F.}\ \bibnamefont {Liu}}, \ and\ \bibinfo
  {author} {\bibfnamefont {C.~P.}\ \bibnamefont {Sun}},\ }\href {\doibase
  10.1103/PhysRevE.90.062125} {\bibfield  {journal} {\bibinfo  {journal} {Phys.
  Rev. E}\ }\textbf {\bibinfo {volume} {90}},\ \bibinfo {pages} {062125}
  (\bibinfo {year} {2014})}\BibitemShut {NoStop}%
\bibitem [{\citenamefont {Dong}\ \emph {et~al.}(2007)\citenamefont {Dong},
  \citenamefont {Yang}, \citenamefont {Liu},\ and\ \citenamefont
  {Sun}}]{dong_quantum_2007}%
  \BibitemOpen
  \bibfield  {author} {\bibinfo {author} {\bibfnamefont {H.}~\bibnamefont
  {Dong}}, \bibinfo {author} {\bibfnamefont {S.}~\bibnamefont {Yang}}, \bibinfo
  {author} {\bibfnamefont {X.~F.}\ \bibnamefont {Liu}}, \ and\ \bibinfo
  {author} {\bibfnamefont {C.~P.}\ \bibnamefont {Sun}},\ }\href {\doibase
  10.1103/PhysRevA.76.044104} {\bibfield  {journal} {\bibinfo  {journal} {Phys.
  Rev. A}\ }\textbf {\bibinfo {volume} {76}},\ \bibinfo {pages} {044104}
  (\bibinfo {year} {2007})}\BibitemShut {NoStop}%
\bibitem [{\citenamefont {Morozov}\ \emph {et~al.}(2012)\citenamefont
  {Morozov}, \citenamefont {Mathey},\ and\ \citenamefont
  {R{\"o}pke}}]{morozov_decoherence_2012}%
  \BibitemOpen
  \bibfield  {author} {\bibinfo {author} {\bibfnamefont {V.~G.}\ \bibnamefont
  {Morozov}}, \bibinfo {author} {\bibfnamefont {S.}~\bibnamefont {Mathey}}, \
  and\ \bibinfo {author} {\bibfnamefont {G.}~\bibnamefont {R{\"o}pke}},\ }\href
  {\doibase 10.1103/PhysRevA.85.022101} {\bibfield  {journal} {\bibinfo
  {journal} {Phys. Rev. A}\ }\textbf {\bibinfo {volume} {85}},\ \bibinfo
  {pages} {022101} (\bibinfo {year} {2012})}\BibitemShut {NoStop}%
\bibitem [{\citenamefont {Marcantoni}(2017)}]{marcantoni_thermodynamics_2017}%
  \BibitemOpen
  \bibfield  {author} {\bibinfo {author} {\bibfnamefont {S.}~\bibnamefont
  {Marcantoni}},\ }\href {\doibase 10.1088/1742-6596/841/1/012019} {\bibfield
  {journal} {\bibinfo  {journal} {J. Phys.: Conf. Ser.}\ }\textbf {\bibinfo
  {volume} {841}},\ \bibinfo {pages} {012019} (\bibinfo {year}
  {2017})}\BibitemShut {NoStop}%
\bibitem [{\citenamefont {Gardiner}\ and\ \citenamefont
  {Zoller}(2004)}]{gardiner_quantum_2004}%
  \BibitemOpen
  \bibfield  {author} {\bibinfo {author} {\bibfnamefont {C.}~\bibnamefont
  {Gardiner}}\ and\ \bibinfo {author} {\bibfnamefont {P.}~\bibnamefont
  {Zoller}},\ }\href@noop {} {\emph {\bibinfo {title} {Quantum noise}}},\
  Vol.~\bibinfo {volume} {56}\ (\bibinfo  {publisher} {Springer},\ \bibinfo
  {year} {2004})\BibitemShut {NoStop}%
\bibitem [{\citenamefont {Chru{\'s}ci{\'n}ski}\ and\ \citenamefont
  {Kossakowski}(2012)}]{chruscinski_markovianity_2012}%
  \BibitemOpen
  \bibfield  {author} {\bibinfo {author} {\bibfnamefont {D.}~\bibnamefont
  {Chru{\'s}ci{\'n}ski}}\ and\ \bibinfo {author} {\bibfnamefont
  {A.}~\bibnamefont {Kossakowski}},\ }\href {\doibase
  10.1088/0953-4075/45/15/154002} {\bibfield  {journal} {\bibinfo  {journal}
  {J. Phys. B}\ }\textbf {\bibinfo {volume} {45}},\ \bibinfo {pages} {154002}
  (\bibinfo {year} {2012})}\BibitemShut {NoStop}%
\bibitem [{\citenamefont {Scully}\ and\ \citenamefont
  {Zubairy}(1997)}]{scully_quantum_1997}%
  \BibitemOpen
  \bibfield  {author} {\bibinfo {author} {\bibfnamefont {M.~O.}\ \bibnamefont
  {Scully}}\ and\ \bibinfo {author} {\bibfnamefont {M.~S.}\ \bibnamefont
  {Zubairy}},\ }\href@noop {} {\emph {\bibinfo {title} {Quantum optics}}}\
  (\bibinfo  {publisher} {Cambridge university press},\ \bibinfo {year}
  {1997})\BibitemShut {NoStop}%
\bibitem [{\citenamefont {Li}\ \emph {et~al.}(2014)\citenamefont {Li},
  \citenamefont {Yang},\ and\ \citenamefont {Sun}}]{li_long-term_2014}%
  \BibitemOpen
  \bibfield  {author} {\bibinfo {author} {\bibfnamefont {S.-W.}\ \bibnamefont
  {Li}}, \bibinfo {author} {\bibfnamefont {L.-P.}\ \bibnamefont {Yang}}, \ and\
  \bibinfo {author} {\bibfnamefont {C.-P.}\ \bibnamefont {Sun}},\ }\href
  {\doibase 10.1140/epjd/e2014-40659-8} {\bibfield  {journal} {\bibinfo
  {journal} {Eur. Phys. J. D}\ }\textbf {\bibinfo {volume} {68}},\ \bibinfo
  {pages} {45} (\bibinfo {year} {2014})},\ \bibinfo {note}
  {arXiv:1303.1266}\BibitemShut {NoStop}%
\bibitem [{\citenamefont {Li}\ \emph {et~al.}(2015)\citenamefont {Li},
  \citenamefont {Cai},\ and\ \citenamefont {Sun}}]{li_steady_2015}%
  \BibitemOpen
  \bibfield  {author} {\bibinfo {author} {\bibfnamefont {S.-W.}\ \bibnamefont
  {Li}}, \bibinfo {author} {\bibfnamefont {C.~Y.}\ \bibnamefont {Cai}}, \ and\
  \bibinfo {author} {\bibfnamefont {C.~P.}\ \bibnamefont {Sun}},\ }\href
  {\doibase 10.1016/j.aop.2015.05.004} {\bibfield  {journal} {\bibinfo
  {journal} {Ann. Phys.}\ }\textbf {\bibinfo {volume} {360}},\ \bibinfo {pages}
  {19} (\bibinfo {year} {2015})},\ \bibinfo {note} {arXiv:
  1407.4290}\BibitemShut {NoStop}%
\bibitem [{\citenamefont {Li}\ \emph {et~al.}(2016)\citenamefont {Li},
  \citenamefont {Kim},\ and\ \citenamefont
  {Scully}}]{li_non-markovianity_2016}%
  \BibitemOpen
  \bibfield  {author} {\bibinfo {author} {\bibfnamefont {S.-W.}\ \bibnamefont
  {Li}}, \bibinfo {author} {\bibfnamefont {M.~B.}\ \bibnamefont {Kim}}, \ and\
  \bibinfo {author} {\bibfnamefont {M.~O.}\ \bibnamefont {Scully}},\ }\href
  {http://arxiv.org/abs/1604.03091} {\bibfield  {journal} {\bibinfo  {journal}
  {arXiv:1604.03091}\ } (\bibinfo {year} {2016})}\BibitemShut {NoStop}%
\bibitem [{\citenamefont {Agarwal}(1971)}]{agarwal_entropy_1971}%
  \BibitemOpen
  \bibfield  {author} {\bibinfo {author} {\bibfnamefont {G.~S.}\ \bibnamefont
  {Agarwal}},\ }\href {\doibase 10.1103/PhysRevA.3.828} {\bibfield  {journal}
  {\bibinfo  {journal} {Phys. Rev. A}\ }\textbf {\bibinfo {volume} {3}},\
  \bibinfo {pages} {828} (\bibinfo {year} {1971})}\BibitemShut {NoStop}%
\bibitem [{\citenamefont {Braunstein}\ and\ \citenamefont {van
  Loock}(2005)}]{braunstein_quantum_2005}%
  \BibitemOpen
  \bibfield  {author} {\bibinfo {author} {\bibfnamefont {S.~L.}\ \bibnamefont
  {Braunstein}}\ and\ \bibinfo {author} {\bibfnamefont {P.}~\bibnamefont {van
  Loock}},\ }\href {\doibase 10.1103/RevModPhys.77.513} {\bibfield  {journal}
  {\bibinfo  {journal} {Rev. Mod. Phys.}\ }\textbf {\bibinfo {volume} {77}},\
  \bibinfo {pages} {513} (\bibinfo {year} {2005})}\BibitemShut {NoStop}%
\bibitem [{\citenamefont {Genoni}\ \emph {et~al.}(2008)\citenamefont {Genoni},
  \citenamefont {Paris},\ and\ \citenamefont
  {Banaszek}}]{genoni_quantifying_2008}%
  \BibitemOpen
  \bibfield  {author} {\bibinfo {author} {\bibfnamefont {M.~G.}\ \bibnamefont
  {Genoni}}, \bibinfo {author} {\bibfnamefont {M.~G.~A.}\ \bibnamefont
  {Paris}}, \ and\ \bibinfo {author} {\bibfnamefont {K.}~\bibnamefont
  {Banaszek}},\ }\href {\doibase 10.1103/PhysRevA.78.060303} {\bibfield
  {journal} {\bibinfo  {journal} {Phys. Rev. A}\ }\textbf {\bibinfo {volume}
  {78}},\ \bibinfo {pages} {060303} (\bibinfo {year} {2008})}\BibitemShut
  {NoStop}%
\bibitem [{\citenamefont {Aurell}\ and\ \citenamefont
  {Eichhorn}(2015)}]{aurell_von_2015}%
  \BibitemOpen
  \bibfield  {author} {\bibinfo {author} {\bibfnamefont {E.}~\bibnamefont
  {Aurell}}\ and\ \bibinfo {author} {\bibfnamefont {R.}~\bibnamefont
  {Eichhorn}},\ }\href {\doibase 10.1088/1367-2630/17/6/065007} {\bibfield
  {journal} {\bibinfo  {journal} {New J. Phys.}\ }\textbf {\bibinfo {volume}
  {17}},\ \bibinfo {pages} {065007} (\bibinfo {year} {2015})}\BibitemShut
  {NoStop}%
\bibitem [{\citenamefont {Kondepudi}\ and\ \citenamefont
  {Prigogine}(2014)}]{kondepudi_modern_2014}%
  \BibitemOpen
  \bibfield  {author} {\bibinfo {author} {\bibfnamefont {D.}~\bibnamefont
  {Kondepudi}}\ and\ \bibinfo {author} {\bibfnamefont {I.}~\bibnamefont
  {Prigogine}},\ }\href {http://doi.wiley.com/10.1002/9781118698723} {\emph
  {\bibinfo {title} {Modern {Thermodynamics}: {From} {Heat} {Engines} to
  {Dissipative} {Structures}}}}\ (\bibinfo  {publisher} {John Wiley \& Sons,
  Ltd},\ \bibinfo {address} {Chichester, UK},\ \bibinfo {year}
  {2014})\BibitemShut {NoStop}%
\bibitem [{\citenamefont {Santos}\ \emph {et~al.}(2017)\citenamefont {Santos},
  \citenamefont {Landi},\ and\ \citenamefont
  {Paternostro}}]{santos_wigner_2017}%
  \BibitemOpen
  \bibfield  {author} {\bibinfo {author} {\bibfnamefont {J.~P.}\ \bibnamefont
  {Santos}}, \bibinfo {author} {\bibfnamefont {G.~T.}\ \bibnamefont {Landi}}, \
  and\ \bibinfo {author} {\bibfnamefont {M.}~\bibnamefont {Paternostro}},\
  }\href {\doibase 10.1103/PhysRevLett.118.220601} {\bibfield  {journal}
  {\bibinfo  {journal} {Phys. Rev. Lett.}\ }\textbf {\bibinfo {volume} {118}},\
  \bibinfo {pages} {220601} (\bibinfo {year} {2017})}\BibitemShut {NoStop}%
\end{thebibliography}%

\end{document}